\documentclass{aastex}
\usepackage{emulateapj5}

\begin{document}

\newcommand{\kms}{$\rm {km}~\rm s^{-1}$}
\newcommand{\ha}{H$\alpha$}
\newcommand{\HI}{\ion{H}{1}}
\newcommand{\Ropt}{R$_{23.5}$}
\newcommand{\etal}{{\it et al. }}

\submitted{Accepted to the {\it Astronomical Journal}}
\title{Maximum Disk Mass Models for Spiral Galaxies}
\author{Povilas Palunas\altaffilmark{1}$^,$\altaffilmark{2}}
\affil{Laboratory for Astronomy \& Solar Physics, NASA/Goddard Space Flight Center, Greenbelt, MD 20771}
\affil{palunas@gsfc.nasa.gov}
\author{T.B. Williams\altaffilmark{1}}
\affil{Department of Physics and Astronomy, Rutgers, The State
University, Box 0849 Piscataway, NJ 08855-0849}
\affil{williams@physics.rutgers.edu}

\altaffiltext{1}{Visiting Astronomer, Cerro Tololo Inter-American Observatory.
CTIO is operated by AURA, Inc.\ under contract to the National Science
Foundation.}

\altaffiltext{2}{NASA/NRC Resident Research Associate}

\begin{abstract}

We present axisymmetric maximum disk mass models for a sample of 74
spiral galaxies taken from the southern sky Fabry-Perot Tully-Fisher
survey (Schommer \etal 1993).  The sample contains galaxies spanning a
large range of morphologies and having rotation widths from 180~\kms\
to 680~\kms.  For each galaxy we have an $I$-band image and a two
dimensional \ha\ velocity field.  We decompose the disk and bulge by
fitting models directly to the $I$-band image.  This method utilizes
both the distinct surface brightness profiles and shapes of the
projected disk and bulge in the galaxy images.  The luminosity
profiles and rotation curves are derived using consistent centers,
position angles, and inclinations derived from the photometry and
velocity maps.  The distribution of mass is modeled as a sum of disk
and bulge components with distinct, constant mass-to-light ratios. No
dark matter halo is included in the fits.

The models reproduce the overall structure of the rotation curves in
the majority of galaxies, providing good fits to galaxies which
exhibit pronounced structural differences in their surface brightness
profiles. 75\% of galaxies for which the rotation curve is measured to
\Ropt\ or beyond are well fit by a mass-traces-light model for the
entire region within \Ropt.  The models for about 20\% of the
galaxies do not fit well; the failure of most of these models is
traced directly to non-axisymmetric structures, primarily bars but
also strong spiral arms.  The median $I$-band M/L of the disk plus
bulge is $2.4\pm0.9{\rm h_{75}}$ in solar units, consistent with
normal stellar populations.  These results require either that the
mass of dark matter within the optical disk of spiral galaxies is
small, or that its distribution is very precisely coupled to the
distribution of luminous matter.

\end{abstract}

\section{Introduction}

Extended \HI\ rotation curves provide deeply compelling evidence for a
dark matter component that dominates the total mass of spiral
galaxies.  However, the fraction of luminous to dark matter (L/D)
within the optical disk (within the 25 mag/arcsec$^2$ $B$-band, 23.5
mag/arcsec$^2$ $I$-band isophote) is very poorly known. Typically,
extended rotation curves have no distinct features which might reflect
the end of the luminous disk (Bahcall \& Casertano 1985).  As a
consequence, they provide virtually no constraint on L/D within the
optical disk.  Mass models in which the dark component dominates the
inner mass distribution generally fit the rotation curves within the
optical disk as well as models with a negligible dark component
(van~Albada \& Sancisi 1986, Lake \& Feinswog 1989).

The mass and luminosity of spiral galaxies are, however, very strongly
connected.  Spiral galaxies exhibit a very tight, one parameter
relation between rotational velocity and luminosity, the Tully-Fisher
(TF) relation (Tully \& Fisher 1977, see Jacoby \etal 1992 for a
review).  The total mass-to-light ratio (M/L) within the optical disk
is remarkably uniform among all spirals (Rubin 1985, Roberts \& Haynes
1994) including low surface brightness spirals (Sprayberry \etal
1995). Indeed, the typical value of M/L (we derive here (M/L)$_I$ =
2.4$\pm$0.9) is consistent with that expected from normal stellar
populations (Larson \& Tinsley 1978, Bruzual \& Charlot 1993, Worthey
1994), with no dark matter. Optical rotation curves are not, in
general, flat and featureless.  They span a range of shapes from
linearly rising to falling with radius (Rubin 1985) and also display
smaller scale ``bumps and wiggles'' (Freeman 1992).  Mass models
fitted to rotation curves within the optical disk rarely require dark
matter halos to yield good fits (Kalnajs 1983, Kent 1986, Buchhorn
1992).  These models generally reproduce the large scale features and
sometimes reproduce the smaller scale ``bumps and wiggles'' in the
optical rotation curves.

Dynamical arguments suggest that the inner regions of spirals cannot
be dominated by dark matter.  Within the optical radius, a dark matter
halo with an average projected surface mass density greater than that
of the disk would act to suppress common instabilities such as bars
and 2-arm spiral structure (Athanassoula \etal 1987).  The existence
of lopsided modes, which appear in many disk galaxies (Rix \& Zaritsky
1995), may require even lower halo mass densities.  A bulge or thick
disk also act to stabilize the disk, further lowering the allowed dark
matter density. It is possible to contsruct a model disk galaxy with
an almost flat rotation curve which is stable with no dark matter
(Sellwood \& Evans 2000).  This reverses the disk-stability argument
Ostriker \& Peebles (1973), which is often used to support the need
for a massive dark matter halo within the optical radius.  Of course,
spherical halo mass outside the optical radius has no effect on disk
stability.  A stellar bar would also interact with a massive halo
through dynamical friction.  For large dark matter densities this
interaction acts to rapidly slow the pattern rotation speed of the bar
(Weinberg 1985, Debattista \& Sellwood 2000) far below that expected
in real galaxies (Sellwood \& Wilkinson 1993), or measured in any
galaxy (Merrifield \& Kuijken 1995, Gerssen, Kuijken \& Merrifield
1999)

Simulations of galaxy formation through dissipationless collapse of
matter yield strongly triaxial halos (Dubinski \& Carlberg 1991,
Warren \etal 1992) , which produce non--axisymmetric disks if the halo
dominates the inner mass density.  Adding a dissipational gaseous
component to these simulations leads to rounder, but still triaxial
halos (Katz \& Gunn 1991, Dubinski 1994).  The intrinsic ellipticity
and non--circular motions of these disks would produce large scatter
in the TF relation (Franx \& de~Zeeuw 1992) which is not observed.
This limits either the halo triaxiality to less than that predicted by
any of the formation models, or the mass of dark matter within the
optical disk.  The shapes of the gravitational potentials of dark
matter halos for two galaxies have been determined through dynamical
models of a polar ring (Sackett \etal 1994) and the flaring of \HI\
gas (Olling 1996).  Both of these studies suggest that halos have very
flat mass distributions, with axis ratios between 0.1 and 0.3, and
thus resemble a disk more closely than a sphere.

The various evidence discussed above strongly suggest that the
luminous component contains a dynamically significant fraction ($\geq
1/2$) of the mass in the inner regions of spiral galaxies.  However,
most rotation curves (see Casertano \& van~Gorkom 1991) for two
possible counter examples) show no significant change in the
transition between this inner region and the outer, dark matter
dominated region.  The disk and halo must somehow conspire to hide
this transition (van~Albada \& Sancisi 1986).  Moreover, because of
the low scatter in the TF relation and in M/Ls, the disk-halo
conspiracy must act consistently among all spiral galaxies to couple
tightly the evolutionary history and the present structure of luminous
disks and dark halos.  This is extremely surprising, because after the
initial collapse the evolution of the halo and disk proceed on
different scales and through different and complex physical processes
which are linked only through a weak gravitational coupling.

All of the above considerations led van~Albada and Sancisi (1986) to
advance the maximum disk hypothesis.  Under this hypothesis, the mass
of the luminous disk in a spiral galaxy is assumed to be as large as
possible, consistent with the galaxy's rotation curve.  The mass
contribution of the dark matter halo is therefore assumed negligible
in the inner parts of spirals.

A maximal disk does not eliminate the disk-halo conspiracy, in a
sense, it makes it more puzzling because it minimizes the overlap in
the distribution of dark and luminous matter.  However, it does make
the tight correlation between mass and luminosity in the inner parts
of spirals more plausible.

Not all lines of evidence support the maximum disk hypothesis.
Kuijken \& Gilmore (1991) quote a local Galactic surface mass density,
derived from the velocity dispersion of K dwarfs, which is 30\% higher
than that of ``identified'' matter.  Cosmological N-body simulations
of hierarchical universes yield halos which are not well approximated
by isothermal spheres with cores (Dubinski \& Carlberg 1991, Warren
\etal 1992, Navarro \etal 1996).  The mass density profiles of these
simulated halos have slowly changing logarithmic slopes and continue
to rise all the way into the centers.  The central mass concentration
of these simulated halos would seem to be inconsistent with a maximum
disk.

In this work we model the luminous mass distribution for a sample of
74 spiral galaxies.  Our aim is to test the maximum disk hypothesis by
analyzing how well features in the rotation curve are reproduced by
the mass models.  We therefore do not include a dark halo in the fits,
but rather evaluate the quality of the fits under the strict maximum
disk hypothesis.  We extract surface brightness profiles for the disk
and bulge from $I$-band images of normal spiral galaxies and derive
optical \ha\ rotation curves from Fabry-Perot velocity maps with two
spatial dimensions.  The optical radius is defined as the radius to
the extinction corrected 23.5 mag~arcsec$^{-2}$ isophote in $I$.  The
Fabry-Perot maps provide high signal-to-noise rotation curves, and
enable us to average over local kinematic features.  We fit
axisymmetric mass models to optical rotation curves assuming constant,
but distinct, mass-to-light ratios for the disk and bulge.  The two
dimensional information provided by the photometric images and
velocity maps allows us to assess the importance of nonaxisymmetric
features. This will be fully explored in a future paper.

\section{The data/sample}
Our sample consists of 74 field and cluster spirals in the vicinity of
the Hydra-Centaurus cluster. For each galaxy in the sample we have an
$I$-band image and a two dimensional H$\alpha$ velocity map.
Sixty--one galaxies are taken from Schommer \etal (1993) (SBWM).  The
observations and preliminary reductions of the data for these galaxies
are described there.  Observations and reductions for 13 additional
galaxies are presented in this paper. The sample includes galaxies
which are members of the clusters: Antlia(6), Hydra(12),
Centaurus(16), and Klemola 27 (5) (see Table 1).

$I$-band CCD images were taken on March 15, 1994 with the CTIO 0.9m
telescope at the Cassegrain focus. The detector was a TEK 1024 with a
scale of 0.39\arcsec/pixel. The exposure time for each image was 10
minutes. Typical seeing was $\sim$1.5\arcsec. The images were bias
subtracted and flatfielded with twilight skyflats using
IRAF\footnote{IRAF is distributed by NOAO, which is operated by AURA
Inc., under contract to the NSF.}. A large fraction of the SBWM images
were taken with a TI 800$\times$800 CCD which had large, 1.5\%,
flatfield errors. The errors were large scale and primarily near the
edges of the field which made precise sky estimation difficult and
created significant distortions at low surface brightness. An
illumination correction was constructed for the TI images by combining
all 40 $I$-band images from the run.  For each image the galaxy and
stars were removed and the remaining pixels were scaled by the sky
level. The resultant image was smoothed over a scale of 20 pixels. The
corrected images are flat to 0.2\%.

All the images were calibrated using Graham standards (Graham
1982). The instrumental magnitudes of the standards were determined
using DAOPHOT II (Stetson 1987).  DAOPHOT was also used to
automatically locate cosmic rays and stars in the galaxy images. Stars
and cosmic rays were distinguished using the DAOPHOT sharpness
parameter. The cosmic rays were removed by replacing affected pixels
with the biweight of surrounding pixels (the biweight is an robust
estimate of the mode of a distribution which is optimized for small
samples (see Beers \etal 1990). The stars were removed by flagging
pixels in a circular aperture centered on each star.

\ha\ spectroscopy for each galaxy was obtained in April 1993 with the
CTIO 1.5m telescope and the Rutgers Fabry-Perot imaging
spectrophotometer. A TEK 512 CCD with a 1.09\arcsec/pixel scale was
used.  The observations for each galaxy consist of 8 to 15 images with
a 110 \kms\ FWHM bandpass (2.4 \AA\ at \ha), spaced at 1 \AA\
intervals to sample the \ha\ emission line. Wavelengths were
calibrated during the day. Calibration drifts during the night were
monitored by taking exposures of a neon lamp every hour. The maximum
drift rate was 0.1 \AA/hr. The images were bias subtracted and
flatfielded, with dome flats taken near the wavelength of each image,
using IRAF.  Typical seeing was $\sim$1.5\arcsec.

Transparency and instrumental throughput variations were measured by
performing photometry on stars common to all the images for each
galaxy using DAOPHOT. Cosmic rays were removed using the procedure
described above. The images were convolved with a Gaussian to
compensate for variations in seeing. The stellar positions were used
to establish transformations and the images were shifted to a common
coordinate system. The sky in each frame was determined in an annulus
centered on the galaxy.  There was generally no observable wavelength
dependent structure in the the sky over the small 6 \AA\ gradient
across Fabry-Perot field of view at the observed wavelengths.  The
series of images yields, at each pixel, a short segment of the
spectrum around \ha. The spectra have been fitted with Voigt profiles
(Humli\v{c}ek 1979) to yield maps of the velocity, velocity
dispersion, \ha\ intensity, continuum intensity, and their respective
uncertainties. The kinematic data extends to a median of 4 disk scale
lengths or 1.1 \Ropt\ in $I$.  We used stars in each of the images to
find relative astrometric corrections between the photometric and
kinematic images.

\section{Spiral Galaxy Models}
We derive axisymmetric mass models of spiral galaxies assuming that
the radial mass distribution follows the radial luminosity
distribution with constant M/L.  The assumption of constant M/L
requires that extinction and population gradients be small across the
luminous disk, which may not be an adequate representation of real
galaxies.  de~Jong (1995) finds significant color gradients in the
profiles of a large sample of spiral galaxies. His analysis attributes
these gradients primarily to changes in stellar populations, with
younger, more metal poor stars at large radii.  This would tend to
lower the stellar M/L at large radii and, therefore, decrease the
radius at which the ``missing mass'' becomes important.  de~Jong's
models predict that (M/L)$_I$ can change by a factor of 1.5 from the
inner to outer parts of spirals.  However, his models predict a
scatter of $\sim$1.5 mag in the $I$-band Tully-Fisher and a smaller
scatter in color bands redder than $I$. Neither of these predictions
are in agreement with observations.  The large scatter is not observed
and an analysis of the TF relation shows larger scatter both at
visible color bands (Bothun \& Mould 1987) and in the near infra-red
(Bernstein \etal 1994) than at $I$. These issues clearly need to be
resolved, but in the absence of specific predictions for variable M/L
we adopt a constant M/L for our models.

We assume that luminous parts of spiral galaxies are composed of two
principal components: a flat disk and a rounder bulge.  Each component
is characterized by a distinct spatial and kinematic stellar
distribution, and stellar population. In addition, the disk harbors
the cold gas. Our models are based on a two component, disk and bulge,
photometric decomposition. To derive the mass distribution we assume
that each component has a separate, constant M/L. In real galaxies,
the disk and bulge can be further divided into subcomponents such as
the thin and thick disk, and nucleus; these are not considered here.
The thick and thin disks in external galaxies have been distinguished
only in edge-on projections (Burstein 1979) and the nuclei are not
resolved in our data.

Spiral galaxies are approximately axisymmetric; the axisymmetry is
broken by bars and spiral structure.  Our models assume strict
axisymmetry. The surface brightness profiles and the rotation curves
for each galaxy are derived with a fixed and consistent center,
position angle and inclination which we derive from both the
photometry and the kinematic data. The difference in the projected
disk and bulge axis ratios is taken into account in the disk-bulge
decomposition. In this analysis there is no radial dependence of the
geometric parameters. Such a dependence is often implicit in isophotal
or kinematic tilted-ring analysis.

Our goal is to examine how well maximum disk models reproduce the mass
distribution in a large and diverse sample of spiral galaxies.
Individual models with halos could be constructed but would be poorly
constrained by our data.

In sections \ref{secpmod} and \ref{seckmod} we present our photometric
and kinematic models. The 2-D nature of our data set allows us to
derive independent geometric parameters from both the photometric
images and the kinematic maps (SBWM). In section \ref{secgeo} we
combine these results to derive a luminosity distribution and rotation
curve with consistent geometric parameters. In section \ref{secmod} we
present the mass models.

\subsection{Photometric Models} \label{secpmod}
The goal of the photometric models is to separate the disk and bulge
and deduce their radial luminosity distributions. The two primary
features we use to distinguish the disk from the bulge in an image are
the difference in the radial surface brightness profile and the axis
ratio of the isophotes.

The surface brightness profiles of most spiral galaxies are
exponential over some fraction of the disk (Freeman 1970).
Approximately 40\% of spirals, however, deviate strongly from purely
exponential disks. These spirals, designated as type II by Freeman,
have profiles that are flat or slowly rising toward the center and
have a steeper exponential outer profile. The profiles of type I
galaxies more closely follow the canonical exponential disk.  The
profiles of disk galaxies often also exhibit smaller scale deviations.
In order to model fully the radial structure of the disk these
features must be included. We use an exponential component for the
disk in our disk-bulge decomposition, but the final model for the disk
is taken as the radially binned surface brightness profile of the
galaxy after subtracting the bulge model.

The bulges of spiral galaxies are three-dimensional structures.  This
leads to the difference in axis ratios of the disk and bulge isophotes
in the projected image of an inclined spiral. The isophotes of the
bulge are typically not circular, and thus bulges are typically not
spherical.

\subsubsection{Disk-Bulge Decomposition}
There have been two strategies for performing disk-bulge
decompositions. One has been to first perform an isophotal analysis;
the resulting radial luminosity profile is fitted with an exponential
disk and a bulge model (Schombert \& Bothun 1987) such as the $R^{1
\over 4}$~ law (de~Vaucouleurs 1948) or Plummer model. The second
method, due to Kent (1986), uses the geometric properties of the disk
and bulge.

The profiles derived from isophotal fits are functions of up to five
variables at each isophote: semi-major axis, ellipticity, position
angle and center coordinates. The dependence on the last four
variables, while included in the fits, is often not presented. This
dependence leads to distortions in the radial luminosity profile which
are not in the underlying radial light distribution. The geometric
parameters can be distorted locally by bright star forming regions.
Unavoidably, the the parameters have a radial dependence that follows
a bar and/or spiral arms. The isophotes are not well approximated by
ellipses in the region where the surface brightness changes from bulge
dominated to disk dominated.  An elliptical isophote fitter makes a
compromise fit which biases the true radial profile. For example, in a
highly inclined galaxy with a compact nearly spherical bulge, the
surface brightness of the bulge projects to large galactic radii along
the minor axis. An elliptical isophote fitter will produce a profile
in which the bulge seems to extend to larger radii. This will bias
bulge models such as the $R^{1 \over 4}$ or Plummer model which have
strong tails.  Byun \& Freeman (1995) present a systematic study of
these effects in model galaxies.

The method of disk-bulge decomposition developed by Kent uses the
different radial scaling properties of the disk and bulge profiles
along the major and minor axes. The method relies on the different
axis ratios of disk and bulge isophotes and the assumption of axial
symmetry.  The disadvantage of this method is that it uses information
only near the major and minor axis and is, therefore, sensitive to
non-axisymmetric structure in these regions.

Our method combines the best properties of these two methods by using
the full two-dimensional information in the galaxy image.  We fit
two-components, a disk and a bulge model, to the image of the galaxy.
We assume that the underlying distribution of light in spiral galaxies
is axisymmetric, with a thin disk and an oblate spheroidal bulge with
constant axis ratio. Under this assumption the projected isophotes of
the model disk are ellipses with constant axis ratio, position angle
and center. The projected isophotes of the model bulge are ellipses
with the same position angle and center as the disk but a different,
larger, axis ratio. We exploit the exponential form of the disk but do
not constrain ourselves to a more specific functional form for the
bulge. For the bulge we use a series expansion of Gaussians
(Bendinelli 1991) which we have generalized to model oblate
distributions. At large radii the surface brightness of this bulge
model is always negligible compared to the disk.

The disk and bulge are separated in three steps: (1) an exponential
disk model is fitted to and subtracted from the image, (2) the bulge
model is fitted to the resultant image, and (3) the bulge model is
subtracted from the original image leaving an image of the disk. The
exponential disk of step 1 is fitted in a region well away from the
bulge and extrapolated into the central bulge dominated region. The
extrapolation is carried out in different ways for type I and type II
disks (see below). The purpose of fitting the exponential disk model
is to derive a global inclination, position angle, and scale length
and to provide a reasonable extrapolation of the disk into the central
bulge dominated region for the purpose of isolating the bulge light.

For type I disks the galaxy image is divided into two regions; the
disk dominated region and the central bulge dominated region. The disk
dominated region is defined as an elliptical annulus centered on the
galaxy: the inner edge of this annulus has axis ratio of the bulge
isophotes and the outer edge has the axis ratio of the disk
isophotes. The semi-major axis of the inner edge of the annulus is set
sufficiently far from the center that the bulge is negligible compared
to the exponential disk.  The bulge dominated region is defined as an
elliptical disk centered on the galaxy with the axis ratio of the
bulge isophotes. The semi-major axis of each division is determined
from a plot of the surface brightness binned in elliptical annuli with
the approximate axis ratio of the disk.  We fit a projected
exponential disk to the image of the galaxy in the disk dominated
region:
\begin{equation}
\mu_{_D}(a_{_D}) = \mu_{_{D_\circ}} e^{- {a_{_D}}/{r_{_D}}},
\end{equation}
where $\mu_{_{D_\circ}}$ is central surface brightness, $r_{_D}$ is
the disk scale length, and  $a_{_D}$ is the length of the semi-major axis of
a ellipse of constant surface brightness for the disk. In general
\begin{equation}
a^2=R^2(1+{\mathrm f}^2{\mathrm sin}^2(\phi - \phi_\circ))
\end{equation}
where ${\mathrm f}^2 = ({\mathrm b \over a})^{-2} - 1$ is defined in
terms of the axis ratio, ${\mathrm b \over a}$, $R$ is the distance
from the galaxy center as measured on the sky and $\phi_\circ$ is the
position angle of the major axis. For a flat disk the inclination is
given by $i = \cos^{-1}({\mathrm b \over a})$.

The exponential disk is subtracted from the galaxy image and a bulge
model is fitted in the bulge dominated region. The bulge model is a
series of Gaussians:
\begin{equation}
\mu_B(a_{_B}) = {\sum^{n}_{k=1}}
{{c_k}\over{\pi r_{_{B_k}}^2}}e^{-{ a_{_B}^2 / r_{_{B_k}}^2}},
\end{equation}
where $c_k$ is the total light in each component, $r_{_{B_k}}$ is the
scale length of the $k^{th}$ Gaussian component and $a_{_B}$ is the
length of the semi-major axis of an ellipse of constant surface
brightness for the bulge.  Each Gaussian has a common center and axis
ratio. The position angle is fixed at the value derived for the disk,
as required by the assumption of axisymmetry. Bulges are fitted with
between 1 and 6 Gaussian components. We use the maximum number of
components that gives a unique and stable fit.  If we add additional
components we find either that: multiple components converge to the
same scale length, with each of these components contributing a
fraction of the intensity, or that the scale length of an additional
component diverges and the intensity is reduced to the point that the
component contributes a negligible constant offset. We have found that
many of the standard fitting laws for the bulges of spiral galaxies
such as de~Vaucouleurs or Plummer models are inadequate
representations of real bulge images. The bulge model is subtracted
from the original galaxy image, and the residual image is binned in
elliptical annuli to arrive at the final disk model.  To avoid
truncation features and to integrate total magnitudes we extrapolate
the disk model to large radii using the parameters of the exponential
disk.

For type II galaxies we divide the galaxy into 3 regions: the outer
disk region, the inner disk region and the central bulge region.  We
fit an exponential disk in the outer region to derive the ellipticity
and position angle. We then fit an exponential disk in the inner disk
region with its own intensity and scale length, but with the
ellipticity and position angle fixed at the value derived from the
outer disk.  Because the surface brightness is nearly constant in this
region, the shapes of the isophotes are not well defined and the light
distribution contains no information about the position angle or
inclination of the galaxy.  We subtract the inner disk model from the
image and the bulge model is fitted in the central region as described
above. The bulge model is subtracted from the original galaxy image
and the residual image is binned in elliptical annuli to arrive at the
final disk model.  The disk model is extrapolated to large radii using
the parameter of the exponential disk derived in the outer region.

The disk-bulge decomposition is not generally iterated, except to
adjust the border of division between the disk and bulge regions. The
disk model near the center is a reasonable, but arbitrary,
extrapolation.  Attempts to make small improvements in the solution
through iteration depend on the details of the assumed disk
extrapolation. The advantages of our method are that it does not start
with built--in biases of the isophotal surface brightness profile, and
it sets additional, geometric, constraints on the disk--bulge
decomposition and therefore exploits the available 2-D information.

\subsubsection{3-D luminosity distribution}
The three dimensional luminosity distribution of the disk is trivially
related to the projected distribution. Assuming no internal extinction
$\rho_{_D} (r) = \mu_{_{D}}(r)\cos i~ \delta(z)$, where r is the
distance from the center of the galaxy and $\delta(z)$ is the Dirac
delta function. The surface brightness decreases by a factor of $\cos
i$ in the deprojection.

Under our assumptions the two-dimensional elliptical bulge surface
brightness distribution can be uniquely de-projected to the
three-dimensional spheroidal luminosity distribution via Abel's
integral equations (Stark 1977). A two-dimensional elliptical Gaussian
distribution de-projects to a three-dimensional spheroidal Gaussian
distribution.

\begin{equation}
\rho_{_B} ({\tilde{a}_{_B}}) = {\sum^{n}_{k=1}} {c_k \over
\sqrt{\pi^3(1-e_{_B}^2) r_{_B}^3}} e^{- {\tilde{a}_{_B} / r_{_B}}},
\end{equation}
where ${\tilde{a}_{_B}}$ is the deprojected semi-major axis of the
spheroid,
\begin{equation}
{\tilde{a}_{_B}}={ x^2+y^2+{1 \over 1-e_{_B}^2}z^2}.
\end{equation}
The ellipticity of the spheroid,
\begin{equation}
e_{_B} =  {\mathrm {f_{_B}^2 \over 1-f_{_B}^2}}{1 \over \sin^2i},
\end{equation}
follows directly from the galaxy inclination and the axis ratio of the
bulge projected on the sky.

\subsubsection{Photometric Corrections}

The luminosity profiles are corrected for internal extinction,
$A_{int}$, and Galactic extinction, $A_{ext}$.  Corrections for
internal extinction by dust generally give the total fraction of light
absorbed within the galaxy.  The surface brightnesses of galaxies in
this sample are derived by assuming the extinction is uniform over the
luminous disk. The problem is, however, considerably more complex. The
actual distribution of dust in galaxies is not well known and
scattering may be as important as absorption in the I-band.
Furthermore, multi-color photometry of spiral galaxies by de~Jong
(1995) suggests that radial color gradients may be due primarily to
population gradients.

We adopt $A_{int} = -1.0{\mathrm log}({b \over a})$ given by
Giovanelli \etal (1994). Similar results are found by Han (1992),
Bernstein \etal (1994), and Willick \etal (1995).  Galactic
extinction, $A_{ext}$ in the $B$-band, is taken from Burstein \&
Heiles (1978).  Reddening between $B$ and $I$ is assumed to be
E(B-I)=0.45.  The median value of Galactic extinction for this sample
is 0.12 mag.

\subsection{Kinematic Models} \label{seckmod}
We derive rotation curves from two-dimensional Fabry-Perot \ha\ radial
velocity fields. We assume that the \ha-emitting gas is in an
axisymmetric rotating thin disk.  In polar coordinates the model,
projected on the sky, is given by

\begin{equation}
v(r,\phi) = v_{sys} + 
v_{circ}(r)\sin{i}{\left[ \cos{i}{\cos(\phi-\phi_\circ)} \over 
\sqrt{1-\sin^2{i}\cos^2(\phi-\phi_\circ)}\right]}
\label{eq_velmod}
\end{equation}
where $i$ is the inclination, $\phi_\circ$ is the position angle of
the projected major axis, $v_{circ}(r)$ is the circular velocity
profile and $v_{sys}$ is the systemic velocity. The disk center is an
implicit pair of parameters in the model. The term in brackets is
equal to the cosine of azimuthal angle in the plane of the galaxy
measured from the major axis.  

The parameters of the kinematic model are derived by fitting to the
2-D data in concentric elliptical annuli using a Levenburg-Marquardt
$\chi^2$ minimization technique (Press \etal 1992).  The covariance
matrix at the $\chi^2$ minimum is used to estimate the errors in the
parameters. The errors in the kinematic center are generally larger
that those of the photometric center. The primary reason for this is
that the kinematic center is poorly constrained along the minor axis
and couples to the systemic velocity along the major axis. The center
was therefore fixed by the centroid of the continuum distribution. In
each annulus we fitted $v_{circ}$, $\phi_\circ$ and $i$.  The global
kinematic position angle and inclination are the average, weighted by
the estimated errors, of these parameters from each ellipsoid.  The
final rotation curve is extracted with all of the geometric parameters
fixed. The rotation velocity is estimated independently on each side
of the minor axis.

\subsection{Geometric Parameters}\label{secgeo}
The photometric and kinematic models yield independent estimates of
the major axis position angle and the inclination. We merge these
results and rederive the models using consistent parameters.

The position angle is generally better constrained by the kinematic
models.  A distinct line of nodes delineates the position angle in the
velocity map, while the photometric position angle depends on the
average distribution of luminosity around an annulus. The photometric
position angle is therefore more easily biased by global
non-axisymmetric features such as spiral arms.

The inclination, however is more poorly constrained by the kinematic
model for many galaxies.  For rising rotation curves the slope of the
rotation curve and the inclination are degenerate parameters.  In the
extreme case of a rigid rotator the degeneracy between the rotation
profile and inclination is complete and there is no independent
kinematic information on the inclination. SBWM found some galaxies
with large deviations between the photometric and kinematic
inclinations. The deviations are primarily for galaxies with rising
rotation curves.

We therefore determine the position angle from the kinematic model and
the inclination from the photometric model.  The models are then
iterated with the position angle fixed in the photometric model and
the inclination fixed in the kinematic model.  The final values of the
geometric parameters are fixed and used consistently in both the
photometric and kinematic models. The values are tabulated in Table 1.

\subsection{Galaxy parameters}
In Table 1 we list the photometric and kinematic parameters for
galaxies in the sample. After fixing the geometric parameters of the
models, we derive values of the central surface brightness
($\Sigma_\circ$) and scale length (${\rm r_d}$) for the disk.  Dust
extinction and projection effects are included in the models.  We
quote ${\rm r_d}$ in the outer parts of the disk.  For Freeman type II
galaxies, therefore, the central surface brightness and scale length
are not directly related quantities. The radius of the 23.5
mag~arcsec$^{-2}$ isophote (\Ropt) is measured directly from the
photometric profiles and, therefore, is also corrected for dust
extinction (assuming the dust acts uniformly over the disk) and
projection effects. The magnitudes for each galaxy are derived by
integrating the disk profile and adding the total bulge luminosity. We
quote total magnitudes integrated to infinity.  We calculate the ratio
of the total luminosity bulge to that of the disk (B/D).

We define the velocity width, used in the Tully-Fisher relation, to be
twice the rotation speed, measured by a weighted average of the
rotation curve points where the rotation curve becomes flat.  For
rotation curves which are still rising we define the width to be twice
the maximum rotation speed.

\subsection{Mass Models}
We assume that the mass distribution of a spiral galaxy follows the
de-projected luminosity distribution with constant M/Ls for each
component. For the disk, we use a Fourier transform method for
computing the rotation curve of a flat axisymmetric mass distribution
given by Kalnajs (1965):

\begin{equation}
v_{circ}^2(u)={1\over{2\pi}} {\int^{+\infty}_{-\infty}} dpB(p)e^{jpu}
\quad,
\end{equation}
where $u={\mathrm ln}r$ and
\begin{equation}
B(p) =  2\pi GA(p)
{{\Gamma( {{1+jp}\over{2}} ) \Gamma(1- {{jp}\over{2}} )}
\over
{\Gamma({{1-jp}\over{2}})\Gamma(1+ {{jp}\over{2}} )}}
\quad {\mathrm and},
\end{equation}
\begin{equation}
A(p) =
{\int^{+\infty}_{-\infty}} due^u\mu(u)e^{-jpu}
\end{equation}
where $\Gamma$ is the Gamma function and $j=\sqrt{-1}$.

For a spheroidal mass distribution the rotation curve in the symmetry
plane is given by Binney \& Tremaine (1987).
\begin{equation}
v_{circ}^2(r,z=0) =
4\pi G\sqrt{1-e_{_B}^2}{\int^r_0} {{\rho(\tilde{a}^2)\tilde{a}^2d\tilde{a}} \over \sqrt{r^2-e_{_B}^2\tilde{a}^2}}
\end{equation}
For a Gaussian distribution this integral reduces to a degenerate
hypergeometric series in two variables.

No dark halo is included.  We also do not include a gas component.
\HI\ may contribute approximately 10\%\ of the mass within the optical
radius of late type spirals (Broeils \& van~Woerden 1994) and thus
further lower the allowed mass of the luminous component. \HI\ gas
disks in spiral galaxies have longer scale lengths than the stellar
disks and therefore contribute to the mass distribution primarily at
larger radii where the rotation velocity due to the stellar disk
begins to fall off.

\section{Results}
The model rotation curves of the bulge and disk are fitted to the
kinematic data by adjusting the M/L of each component; the best fit is
derived by minimizing $\chi^2$.  For most galaxies this results in a
fit that nowhere significantly exceeds the data. For galaxies which
would require a halo to get an acceptable fit we limit the radius of
the fit so that the model rotation curve does not exceed the rotation
curve data and evaluate the quality of the fit out to this radius.

In Fig.\ \ref{fig_massmod} we present the models.  The upper panel for
each galaxy shows the face-on surface brightness profile.  The
contribution of the bulge is indicated by the dashed lines.  The
central surface brightness of the disk is marked with a diamond.  The
lower panel shows the rotation curve and the fitted mass model.  The
rotation speed for the side receding with respect to the center is
marked with crosses and that for the the approaching side is marked
with open circles.  The rotation curves for the model bulge and disk
are traced with dashed lines.  The full model rotation curve is equal
to the disk and bulge models summed in quadrature and is indicated by
an unbroken line.  In addition, the radii of prominent features, such
as bars or rings, are marked with a vertical dot-dashed line. A scale
bar in arcsec and the inclination are given to gauge the effect of
seeing.


\subsection{Morphology of Surface Brightness Profiles}

In addition to the relative brightness of the disk and bulge, the
surface brightness profiles of spirals can be classified by the
morphology of their disks.  Two common assumptions are that spiral
galaxies have an exponential disk and that they have a universal
constant central surface brightness (Freeman's Law 1970).  The
discovery of low surface brightness galaxies (Bothun \etal 1991,
Schombert \etal 1992, McGaugh \etal 1995) most clearly indicates that
spiral galaxies fail the second assumption. The surface brightness
profiles of disks are also not strictly exponential.  The most
prominent distinction in disk profiles is between Freeman types I and
II, described in section~\ref{secpmod} Galaxy disks also frequently
exhibit less extreme deviations from a pure exponential, such as a
point of inflection where the scale length changes.

One statement of Freeman's Law is that the ratio of the optical disk
radius to the exponential scale length, \Ropt/r$_{\rm d}$, is a
constant. In Fig.\ \ref{fig_csbR} we plot \Ropt/r$_{\rm d}$ vs. the
central surface brightness. The tight correlation for exponential
(i.e. type I) disks follows trivially from the definition of the
optical radius and the the wide distribution of central surface
brightnesses in our sample. More significantly, \Ropt/r$_{\rm d}$
correlates with absolute magnitude (Fig.\ \ref{fig_MR}).  We test the
significance of this correlation with the Spearman rank-order
correlation coefficient (Press \etal 1992).  The correlation
coefficient is $-0.29$. The probability that there is no correlation
is 0.01.  The low luminosity galaxies have lower central surface
brightness and relatively flatter surface brightness profiles over the
optical disk as indicated by a larger \Ropt/r$_{\rm d}$. This leads to
model disk rotation curves which rise more slowly for low luminosity
galaxies than those for high luminosity galaxies.


\subsection{Morphology of Rotation Curves}

Within the optical radius spiral galaxy rotation curves are not
generally flat, but span a range of morphologies, from rising linearly
to falling with radius (Rubin 1985, Persic \& Salucci 1995).  These
authors also show that the shape of the rotation curves correlate with
luminosity; low luminosity, small rotation velocity galaxies have
rotation curves that are rising while high luminosity, large rotation
velocity galaxies have falling rotation curves.

An examination of the rotation curves in this sample suggests such a
trend, but a large range of rotation curve shapes is found at every
scale of rotational velocity.  At high maximum rotation velocities
($\sim$300 \kms): ESO~374G02 (Fig.\ \ref{fig_massmod}z) is falling,
ESO~375G12 (Fig.\ \ref{fig_massmod}ac) is flat, ESO~269G61 (Fig.\
\ref{fig_massmod}i) rises over $\sim$0.4\Ropt\ before flattening, and
ESO~381G51 (Fig.\ \ref{fig_massmod}ai) rises over the entire optical
radius. At low maximum rotation velocities ($\sim$100 \kms):
ESO~374G03 (Fig.\ \ref{fig_massmod}aa) is flat past $\sim$0.25\Ropt,
ESO~322G19 (Fig.\ \ref{fig_massmod}k) is flat past $\sim$0.5\Ropt,
Abell~1644d83 (Fig.\ \ref{fig_massmod}a) rises over the entire optical
radius. The rotation curve of ESO~441G21 (Fig.\ \ref{fig_massmod}ba)
also rises to about 100 \kms over the optical radius but does so
almost linearly.

\subsection{Model Fits}\label{secmod}

In spite of the variety of surface brightness profile and rotation
curve morphologies, maximum disk models with constant M/Ls provide
good fits to optical rotation curves for a majority of the galaxies in
our sample.  The overall structure of the rotation curves is
reproduced by the models.  The models fit the data for rotation curves
that are rising linearly or with a curve, that are flat or falling, or
that have strong inflection points.  The galaxies with good fits span
a range of velocity widths (2$v_{\rm circ}$) from 180 \kms\ for
Abell~1644d83 (Fig.\ \ref{fig_massmod}a) to 680 \kms\ for ESO~572G17
(Fig.\ \ref{fig_massmod}bv).  For most of the galaxies no halo is
required for the models to fit the data within the optical radius or
to the last measured point.

Fig.\ \ref{fig_fithist} shows the distribution of the ratio of the
maximum radius out to which the mass models provide a good fit to the
optical radius; R$_{fit}$/\Ropt. The histogram is shaded for models
which provide a good fit out to the last measured point of the
rotation curve.  75\% of galaxies for which the rotation curve is
measured to \Ropt\ or beyond are well fit by a mass-traces-light model
for the entire region within \Ropt.  For 21\% of the galaxies the
models provide poor fits due to strong bars or spiral arms, these
cases are assigned R$_{fit}$/\Ropt = 0 in the histogram. The existence
of strong non-axisymmetric structures suggests that there should be
less dark matter in these galaxies. These cases are discussed below in
section \ref{secnonax}.


For Freeman type I spiral galaxies, a thin exponential disk mass
distribution has a rotation curve which reaches maximum at 2.15 disk
scale lengths and falls slowly thereafter, dropping 10\% by 3.75 disk
scale-lengths. Despite the apparent restriction of this shape, a wide
range of optical rotation curve shapes can be successfully modeled
because of the variation of the number of disk scale lengths within
the optical radius among galaxies and the addition of a bulge
component.

Few galaxies have the canonical flat rotation curve across the entire
optical disk.  ESO~375G12 (Fig.\ \ref{fig_massmod}ac) and ESO~376G02
(Fig.\ \ref{fig_massmod}ae) are two examples of galaxies with good
fits that do.  The optical radii in these galaxies span 4.2 and 3.8
disk scale-lengths, respectively. The model rotation curve for
ESO~375G12 begins to fall at the optical radius, but unfortunately,
the rotation curve data for ESO~375G12 extend to only 3.3 disk scale
lengths.  ESO~376G02 shows slight evidence of dark matter, but only
near the optical radius.  The rotation curve for ESO~374G02 (Fig.\
\ref{fig_massmod}z) has a gentle linear falloff over most of the
optical radius and the model provides an excellent fit over this
entire range. The optical radius of ESO~374G02 extends over 4.1 disk
scale lengths.  These galaxies have prominent bulges.  The flatness of
the rotation curves is achieved though a combination of the bulge and
disk rotation curves.

ESO~383G88 (Fig.\ \ref{fig_massmod}am) and ESO~323g42 (Fig.\
\ref{fig_massmod}x) have less prominent bulges.  The rotation curves
of these galaxies rise less dramatically than those for the flat
rotation curves above, but do flatten at larger radii.  The optical
radius of these spirals extend to 3.2 and 3.7 disk scale-lengths,
respectively, and the models reproduce the turnover in the rotation
curves.

The optical disks of ESO~376G10 (Fig.\ \ref{fig_massmod}af) and ESO
501g01 (Fig.\ \ref{fig_massmod}bl) extend to only 2.0 and 2.2 disk
scale-lengths, respectively.  These galaxies also have small bulges.
The rotation curves reflect this morphology; they rise with a curve
over most of the optical disk, reaching maximum near the optical
radius.  The rotation curve for ESO~501g01 extends significantly past
the optical disk to 4.1 scale lengths.  The model rotation curve does
not fall significantly below the data over this entire range.

Freeman type II disks, which are distinguished by a flat inner surface
brightness profile, are modeled with a constant inner mass density.
This leads to rotation curves which are linearly rising over the
constant density region. The radius of the turnover varies from 20 to
60\% of the optical radius for the galaxies in our sample. Examples
include: ESO~445G81, 0.22\Ropt\ (Fig.\ \ref{fig_massmod}bi); ESO
375G02, 0.23\Ropt\ (Fig.\ \ref{fig_massmod}ae); ESO~216G20, 0.31\Ropt\
(Fig.\ \ref{fig_massmod}c); ESO~501G86, 0.55\Ropt\ (Fig.\
\ref{fig_massmod}bp); and ESO~509G91, 0.62\Ropt\ (Fig.\
\ref{fig_massmod}bs).  The size of the linearly rising region in the
rotation curve varies accordingly.

ESO~381G51 (Fig.\ \ref{fig_massmod}ai), ESO~435G26 (Fig.\
\ref{fig_massmod}an), ESO~438G15 (Fig.\ \ref{fig_massmod}aw) and
ESO~501g11 (Fig.\ \ref{fig_massmod}bm) also have flat surface
brightness profiles near their centers; however in these cases the
profiles roll off slowly approaching an exponential asymptotically
near the edge of the optical disk.  The rotation curves in these cases
rise slowly, with a curve matched by those of the models.  ESO~435G26
has a strong bar and the fit within the bar radius is not good.

The profiles of some spirals have inflections which are less
prominent: the slope of the profile changes but does not become flat.
This feature is reflected in the rotation curves of ESO~317G41 (Fig.\
\ref{fig_massmod}j) and ESO~322G82 (Fig.\ \ref{fig_massmod}s).

\subsection{Nonaxisymmetric Structure}\label{secnonax}
A fraction, about 20\%, of the fits fail in the inner regions; major
structures in the model and/or rotation curve do not match up.  These
bad fits occur well within the optical disk and are not likely due to
a dominant dark matter component.  The galaxies which have the poorest
fitting mass models often have strong non-axisymmetric structures in
the form of bars or strong spiral arm structure.  These structures
affect both the surface brightness profile and the rotation curve.
The strong non-axisymmetric gas motions induced by a bar distort the
measured rotation curve within the bar radius as noted for ESO~435G26.
Strong spiral structure can affect the shape of the surface brightness
profiles and induce large non-axisymmetric motions as well as bias the
determination of inclination and major-axis position angle.

ESO~268g37 (Fig.\ \ref{fig_massmod}g) and ESO~374g03 (Fig.\
\ref{fig_massmod}aa) both have bars along the major axis and very
strong spiral structure.  ESO~323g39 (Fig.\ \ref{fig_massmod}w) has a
bar along the minor axis and strong spiral ams.  ESO~323g25 (Fig.\ 1u)
has very strong grand design spiral arms.  The inclination of this
galaxy, combined with the pitch angle and position angle of the spiral
arms conspire in such a way that the arms closely follow the
ellipticity of the disk from 5 to 10 kpc.  The large structure seen in
the model at these radii is due to this chance alignment.  This effect
is also seen in ESO~322g36 (Fig.\ \ref{fig_massmod}l) and ESO~569G17
(Fig.\ \ref{fig_massmod}bu) which also have strong spiral structure.

A detailed analysis of the effect of strong non-axisymmetric
structures will be presented in future paper.  However, if a dark
matter halo dominates the mass within the optical radius of these
galaxies, it should act to stabilize the disk against these
non-axisymmetric modes.  It is difficult, therefore, to attribute the
poor fit of these models to a large dark-to-luminous mass fraction.

\subsection{Small-Scale Structure}

Mass models for spiral galaxies sometimes, but not always, reveal a
correlation between small scale ``bumps and wiggles'' in the surface
brightness profile and the rotation curve (Kent 1986, Freeman 1992).
We find that in the two dimensional maps the residuals in the
photometry and the residual kinematic motions are highly correlated.
How these structures show up in one dimensional, radial, surface
brightness profiles and rotation curves depends strongly on how the
data is sampled.  The ``bumps and wiggles'' correlation is therefore
most probably due to local perturbations, such as spiral arms and
spiral arm streaming motions, rather than the global mass
distribution.

\subsection{Mass-to-Light ratios}  \label{secml}

The median $I$-band M/L (disk plus bulge) for our sample of galaxies
is $2.4{\rm h_{75}}$ in solar units with an rms scatter of 0.9.  This
accords well with the M/L predicted from stellar population synthesis
models.  If we exclude the galaxies with the worst fits the median M/L
rises to $2.7\pm0.8$. Worthey (1994) estimates that a normal stellar
population which forms in a single burst will have an initial $I$-band
M/L of about 1 in solar units and will increase to about 5 over a time
of 15 Gyr.  His models have an M/L of 2.4 at an age of 6 Gyr, and an
M/L of 2.7 at 8 Gyr.  Spiral galaxies are composed of stellar
populations that span a range of ages and their M/L depends on the
number of stars formed in each generation.  A typical spiral might be
about 10 Gyr old and form most of its stars over the first 4 Gyr.

The theoretical M/Ls are highly dependent on the assumed mass function
for the stellar population and the details of the star formation
history.  Thus the agreement of our M/Ls with theory, although
noteworthy, is not in itself overwhelming evidence for the maximum
disk hypothesis.  The light comes primarily from a small population of
bright, high mass stars, while the most of the mass is in a large
population of faint, low mass stars.

Current star formation rates vary with Hubble type, with lower rates
in early type spirals.  M/Ls increase over time as the stellar
population fades, and therefore M/Ls should also vary with type, with
higher M/Ls in early types. Such a correlation was found by Rubin
(1985) and Kent (1986) for M/Ls in the $V$-band.  In the $I$-band this
correlation is expected to be considerably weaker, and indeed it is
not strong in our data (Fig.\ref{fig_mltype}).  The stellar population
models predict a rate of evolution of M/Ls in $I$-band that is a
factor of two less than that in $V$-band for a given generation of
stars.  Also, the scatter in our M/Ls for a given type is large.


The plot of M/L versus axial ratio (Fig.\ \ref{fig_mlax}) is flat
which gives us confidence that the assumed internal extinction law is
correct and confirms the work of Han (1992), Giovanelli \etal (1994),
Bernstein \etal (1994) and Willick \etal (1995).  The good agreement
of the values of M/L with the stellar population models also indicate
that internal extinction cannot be grossly higher, as claimed by
Valentijn (1990).


The scatter in the M/Ls in our sample is 37\%. One of the primary
sources of error is uncertainty in the distance; measured M/Ls are
inversely proportional to the assumed distance.  The sample is
concentrated in the great attractor region (Dressler \etal 1987),
which is dominated by the Hydra-Centaurus supercluster.  The clusters
are likely to have large peculiar velocities.  Galaxies in the
vicinity of the Centaurus cluster have a median M/L of $1.8\pm.6$ and
galaxies near Hydra have a median M/L of $3.4\pm1$.  The Centaurus
cluster is composed of two major sub-clusters (Lucey \etal 1986) and
the Hydra cluster is also thought to have substructure (Fitchett \&
Merritt 1988).  The galaxies which are members of these clusters may
therefore have especially large peculiar velocities.  Fig.\
\ref{fig_mlhist} gives the distribution of M/Ls for the sample.
Smooth Hubble flow is, unfortunately, not a precise approximation for
estimating the distances to the galaxies in this sample.  The large
scale peculiar motions may contribute 25\% or more to the error budget
(Bothun \etal 1992, Mathewson \etal 1992, da Costa \etal 1996).  This
can also be seen in the TF relation for this sample.


Figure \ref{fig_TF} shows the TF relation for the galaxies in our
sample assuming Hubble flow distances. We plot the total magnitude
with closed symbols for Freeman type I and open symbols for type II
galaxies. We also indicate the \Ropt\ isophotal magnitude at the end
of the line connected to each symbol. The length of the line indicates
the extrapolation from the isophotal to total magnitude.  The fainter
galaxies clearly require a larger extrapolation; this is because the
fainter galaxies have lower central surface brightness as noted above.


If the slope of the TF relation is assumed to be near 6 and we
consider the total magnitudes (Mathewson 1992, Bernstein \etal 1994)
then the scatter around the TF line is 0.46 mag.  The zero point of
the relation is fixed so that the average deviation from the TF line
is zero (this assumes no bulk flow for the sample).  If the deviations
from the TF line are attributed solely to peculiar motions, the error
in the Hubble flow distances has a scatter of 24\%.  If the slope of
the TF relation is assumed to be 10 and we consider isophotal
magnitudes (SBWM, Peletier \& Willner 1993) then the scatter about the
TF relation is 0.75 mag which implies deviations from the Hubble flow
distances of 40\%.

\section{Conclusions} \label{conclude}

We find that spiral galaxy mass models which assume the maximum disk
hypothesis yield good rotation curve fits within the optical radius,
for a variety of spirals with distinct surface brightness and rotation
curve morphologies.  75\% of galaxies for which the rotation curve is
measured to \Ropt\ or beyond are well fit by a mass-traces-light model
for the entire region within \Ropt. It is particularly striking that
spirals with very different disk surface brightness profiles,
generically distinguished by Freeman types I and II, are well modeled
under the maximum disk hypothesis.

Freeman type II galaxies constitute a significant fraction of spiral
galaxies which fail to meet the canonical assumption that all spirals
have an exponential disk.  This has certainly contributed to the large
discrepancy in disk scale lengths published by different authors for
the same galaxies (Knapen \& van~der~Kruit 1992).  Type II galaxies
can be further characterized by the size of the inner flat region and
turnover rate.  They span a large range of velocity widths and show no
correlation with type. They generally break the correlation of
rotation curve shape to absolute magnitude found by Rubin (1985) and
Persic \& Salucci (1995).

Galaxies for which our models fail to give good fits in the inner
regions, $\sim20$\% of the sample, generally have strong features,
particularly bars but also strong spiral arms, which break the
assumption of axisymmetry.  Smaller scale deviations in the surface
brightness profiles and rotation curves, ``bumps and wiggles'', can
also often be traced to nonaxisymmetric features.

These results show that, within the optical regions of most spiral
galaxies, the radial mass distribution is tightly coupled to the
luminosity distribution.  This is a much stronger constraint than that
due to global correlations such as the TF relation.  It implies that
either the mass of dark matter must be small within the optical radius
or that the distribution of dark matter must be precisely coupled to
the distribution of luminous matter.  A dark halo which is independent
of, and unresponsive to the luminous disk cannot dominate the mass
within the optical radius.  A fraction of the derived luminous mass
could be traded for halo mass, but the luminous mass traces the
overall features of so many and various rotation curves that this
fraction could not reasonably be too large.

Persic and Salucci (1995) have constructed synthetic rotation curves
by averaging over 500 optical rotation curves from Mathewson \etal
(1992). They fit each synthetic rotation curve with model rotation
curves for an isothermal halo and an exponential disk in which the
scale length is fixed relative to the optical radius. They conclude
that spirals galaxies with rotation velocities of 150 \kms\ are over
40\% dark matter within the optical radius, and that galaxies with
rotation velocities of 100 \kms\ are over 75\% dark matter.  The
assumptions which lead to this conclusion are that all spirals
galaxies have the same \Ropt/r$_{\rm d}$ and that all spirals of a
given luminosity have the same rotation curve shape.  Our results
strongly suggest that both of these assumptions are too simplifying and
call their results into question.  We find that \Ropt/r$_{\rm d}$ is
larger in low rotation velocity galaxies.  This leads to more slowly
rising rotation curves.  Relaxing only this assumption considerably
weakens their dark matter constraints.


Cosmological N-body simulations of hierarchical universes suggest a
universal halo profile for mass scales from ranging from dwarf
galaxies to rich clusters of galaxies (Navarro \etal 1996).  The halo
profiles are centrally concentrated and for spiral galaxies they
dominate the mass distribution at all radii. In the cold-dark matter
models of Navarro \etal, a galaxy with a maximum rotation velocity of
300 \kms\ is 72\% dark matter within the optical radius, a galaxy with
a maximum rotation velocity of 200 \kms\ is 90\% dark matter, and a
galaxy with a maximum rotation velocity of 100 \kms\ is 96\% dark
matter.  Our results show that optical rotation curves in real
galaxies exhibit a variety of shapes and that these shapes are well
modeled by the luminous distribution of matter.  It is difficult to
reconcile our results with a universal rotation curve in which the
central attraction primarily due to a dark matter halo.

The success of the maximum disk hypothesis in modeling the mass
distribution in the inner parts of spiral galaxies implies that either
the mass of dark matter has to be small or that its projected
distribution must follow precisely that of the luminous matter out to
nearly the optical radius.

\acknowledgments We thank Ben Weiner, Jerry Sellwood and Liz Moore for
critical readings of this paper.  We also warmly acknowledge the
excellent support of the CTIO observing staff. This research has made
use of the NASA/IPAC Extragalactic Database (Ned) which is operated by
the Jet Propulsion Laboratory, Caltech, under contract with the
National Aeronautics and Space Administration.  The Rutgers
Fabry-Perot instrument was built with support from Rutgers University
and from the National Science Foundation grant AST 83-19344.  The RFP
is operated at CTIO under a cooperative agreement between Rutgers
University and the Cerro Tololo Interamerican Observatory.

\clearpage

\begin{deluxetable} {llr|ccrcrrr|rr|rrrr}
\rotate
\tablecolumns{16}

\tablewidth{619.62625pt}
\tablecaption{}

\tablehead{
\colhead{galaxy} &
\colhead{clust} &
\colhead{D} &
\colhead{Htype} &
\colhead{Ftype} &
\colhead{m$_I$} &
\colhead{$\mu_\circ$} &
\colhead{R$_{23.5}$} &
\colhead{r$_{\rm d}$} &
\colhead{B/D} &
\colhead{$i$} &
\colhead{$\phi$} &
\colhead{log(2$v_{\circ}$)} &
\colhead{M/L$_{\rm D}$} &
\colhead{M/L$_{\rm B}$} &
\colhead{M/L} \\
\colhead{} &
\colhead{} &
\colhead{Mpc} &
\colhead{RC3} &
\colhead{} &
\colhead{mag} &
\colhead{mag/${\prime\prime}^2$} &
\colhead{kpc} &
\colhead{kpc} &
\colhead{} &
\colhead{\arcdeg} &
\colhead{\arcdeg} &
\colhead{\kms} &
\colhead{$\odot$} &
\colhead{$\odot$} &
\colhead{$\odot$} \\
\colhead{(1)} &
\colhead{(2)} &
\colhead{(3)} &
\colhead{(4)} &
\colhead{(5)} &
\colhead{(6)} &
\colhead{(7)} &
\colhead{(8)} &
\colhead{(9)} &
\colhead{(10)} &
\colhead{(11)} &
\colhead{(12)} &
\colhead{(13)} &
\colhead{(14)} &
\colhead{(15)} &
\colhead{(16)} 
}

\startdata
a1644d83 (a) &         &  79.62  &    10  &     I &  14.42 & 21.23 &  7.32 & 3.64 & 0.06 &  77   &   278 &  2.29 & 3.00 & 0.15 & 2.84 \\
e215g39  (b) &         &  61.29  &     5  &    II &  12.01 & 19.74 & 12.92 & 4.20 & 0.08 &  50   &    29 &  2.48 & 2.16 & 0.81 & 2.06 \\
e216g20  (c) &         &  77.85  &    10  &    II &  12.07 & 19.35 & 13.46 & 2.90 & 0.09 &  74   &   306 &  2.65 & 2.42 & 0.83 & 2.29 \\
e263g14  (d) &         &  69.83  &     3  &     I &  11.32 & 18.54 & 15.93 & 3.60 & 0.04 &  60   &   289 &  2.54 & 0.92 & 0.63 & 0.91 \\
e267g29  (e) &         &  76.23  &     2  &     I &  11.84 & 19.31 & 15.86 & 4.31 & 0.12 &  51   &   313 &  2.66 & 2.94 & 0.48 & 2.68 \\
e267g30  (f) &         &  75.92  &     3  &     I &  11.48 & 19.77 & 16.40 & 4.98 & 0.62 &  55   &   294 &  2.72 & 4.32 & 1.14 & 3.10 \\
e268g37  (g) &    Cen  &  68.50  &     5  &     I &  12.39 & 19.65 & 12.50 & 3.43 & 0.08 &  55   &   120 &  2.50 & 2.33 & 0.00 & 2.17 \\
e268g44  (h) &    Cen  &  49.95  &     3  &     I &  12.22 & 18.87 &  8.38 & 1.91 & 0.04 &  62   &   244 &  2.48 & 2.04 & 1.54 & 2.02 \\
e269g61  (i) &         &  69.40  &     3  &     I &  10.81 & 19.17 & 24.00 & 5.97 & 0.03 &  76   &   253 &  2.74 & 2.59 & 0.00 & 2.51 \\
e317g41  (j) &         &  81.17  &     2  &     I &  11.53 & 18.89 & 17.66 & 4.19 & 0.03 &  71   &   105 &  2.69 & 2.43 & 1.09 & 2.39 \\
e322g19  (k) &    Cen  &  45.23  &     6  &     I &  12.64 & 19.90 &  8.10 & 2.47 & 0.00 &  79   &   300 &  2.41 & 2.06 & 0.00 & 2.06 \\
e322g36  (l) &    Cen  &  43.51  &     4  &     I &  10.83 & 19.06 & 14.19 & 3.50 & 0.04 &  53   &   104 &  2.50 & 1.28 & 0.22 & 1.24 \\
e322g42  (m) &    Cen  &  55.99  &     5  &    II &  11.91 & 20.76 & 14.57 & 4.65 & 0.05 &  71   &    42 &  2.38 & 1.47 & 0.00 & 1.39 \\
e322g44  (n) &    Cen  &  52.88  &     5  &     I &  11.57 & 19.79 & 12.53 & 3.71 & 0.32 &  68   &    89 &  2.44 & 1.69 & 0.76 & 1.46 \\
e322g45  (o) &    Cen  &  44.19  &     5  &     I &  11.52 & 19.12 & 10.36 & 2.85 & 0.03 &  67   &   308 &  2.52 & 1.73 & 0.00 & 1.69 \\
e322g48  (p) &    Cen  &  60.91  &     3  &     I &  12.59 & 19.94 & 10.56 & 2.87 & 0.00 &  76   &    36 &  2.36 & 0.99 & 0.00 & 0.99 \\
e322g76  (q) &    Cen  &  64.28  &     4  &    II &  11.97 & 20.50 & 12.22 & 2.91 & 0.30 &  57   &   259 &  2.54 & 2.34 & 1.27 & 2.09 \\
e322g77  (r) &    Cen  &  38.19  &     3  &     I &  11.49 & 18.84 &  8.57 & 1.71 & 0.04 &  70   &   172 &  2.61 & 3.61 & 0.57 & 3.49 \\
e322g82  (s) &    Cen  &  65.84  &     5  &     I &  11.05 & 19.46 & 19.16 & 5.34 & 0.11 &  63   &     8 &  2.63 & 2.28 & 1.43 & 2.20 \\
e322g87  (t) &    Cen  &  52.13  &     3  &    II &  11.47 & 19.87 & 14.11 & 4.52 & 0.02 &  80   &   138 &  2.53 & 2.40 & 0.60 & 2.37 \\
e323g25  (u) &    Cen  &  59.76  &     4  &     I &  11.34 & 18.69 & 14.27 & 3.31 & 0.01 &  55   &   283 &  2.66 & 2.61 & 2.75 & 2.61 \\
e323g27  (v) &    Cen  &  54.90  &     5  &    II &  11.14 & 19.98 & 15.85 & 3.76 & 0.05 &  58   &   275 &  2.63 & 2.53 & 2.54 & 2.53 \\
e323g39  (w) &    Cen  &  69.90  &    10  &     I &  13.35 & 20.46 &  9.35 & 3.42 & 0.07 &  53   &   265 &  2.33 & 2.02 & 0.00 & 1.89 \\
e323g42  (x) &    Cen  &  59.73  &    10  &     I &  11.53 & 19.57 & 16.61 & 4.44 & 0.03 &  69   &    79 &  2.45 & 1.71 & 0.00 & 1.66 \\
e323g73  (y) &         &  69.63  &    10  &     I &  12.44 & 18.46 &  9.47 & 2.06 & 0.02 &  48   &   358 &  2.51 & 1.13 & 0.00 & 1.10 \\
e374g02  (z) & Antlia  &  41.42  &     3  &     I &   9.91 & 19.08 & 18.08 & 4.42 & 0.45 &   52  &   303 &  2.71 & 2.63 & 1.00 & 2.12 \\
e374g03 (aa) & Antlia  &  43.22  &     6  &     I &  11.51 & 20.11 & 13.20 & 4.25 & 0.03 &   71  &   149 &  2.34 & 1.45 & 0.00 & 1.41 \\
e375g02 (ab) & Antlia  &  43.75  &     3  &    II &  12.01 & 19.17 &  8.65 & 2.15 & 0.01 &   64  &    20 &  2.45 & 1.73 & 0.13 & 1.72 \\
e375g12 (ac) & Antlia  &  42.86  &     3  &     I &   9.28 & 18.97 & 27.64 & 6.66 & 0.12 &   44  &   309 &  2.75 & 2.19 & 2.18 & 2.19 \\
e375g29 (ad) &         &  56.12  &     5  &     I &  11.99 & 19.83 & 13.15 & 3.54 & 0.00 &   80  &   321 &  2.43 & 1.95 & 0.00 & 1.95 \\
e376g02 (ae) &         &  59.44  &     4  &    II &  11.35 & 19.21 & 14.68 & 3.87 & 0.16 &   75  &   159 &  2.62 & 2.48 & 0.92 & 2.26 \\
e376g10 (af) &         &  46.21  &     8  &     I &  11.00 & 20.85 & 19.53 & 8.12 & 0.02 &   76  &    93 &  2.55 & 3.92 & 0.00 & 3.86 \\
e377g11 (ag) &         &  46.01  &     2  &    II &  10.72 & 19.85 & 17.35 & 3.51 & 0.12 &   73  &    58 &  2.59 & 2.40 & 0.00 & 2.14 \\
e381g05 (ah) &         &  79.56  &    10  &     I &  13.30 & 19.35 &  9.33 & 2.42 & 0.02 &   41  &   299 &  2.49 & 2.25 & 0.00 & 2.19 \\
e381g51 (ai) &         &  70.72  &     3  &    II &  11.44 & 19.34 & 15.18 & 2.71 & 0.12 &   82  &    58 &  2.70 & 2.92 & 0.00 & 2.61 \\
e382g06 (aj) &         &  65.44  &    10  &     I &  13.20 & 19.60 &  8.51 & 2.33 & 0.01 &   54  &    86 &  2.47 & 3.28 & 0.00 & 3.25 \\
e382g58 (ak) &         & 106.20  &     4  &     I &  11.21 & 19.94 & 32.75 & 9.42 & 0.12 &   79  &   153 &  2.80 & 4.32 & 0.79 & 3.93 \\
\enddata
\end{deluxetable}

\begin{deluxetable} {llr|ccrcrrr|rr|rrrr}
\tablenum{1}
\rotate
\tablecolumns{16}

\tablewidth{619.62625pt}
\tablecaption{\it Continued}

\tablehead{
\colhead{galaxy} &
\colhead{clust} &
\colhead{D} &
\colhead{Htype} &
\colhead{Ftype} &
\colhead{m$_I$} &
\colhead{$\mu_\circ$} &
\colhead{R$_{23.5}$} &
\colhead{r$_{\rm d}$} &
\colhead{B/D} &
\colhead{$i$} &
\colhead{$\phi$} &
\colhead{log(2$v_{\circ}$)} &
\colhead{M/L$_{\rm D}$} &
\colhead{M/L$_{\rm B}$} &
\colhead{M/L} \\
\colhead{} &
\colhead{} &
\colhead{Mpc} &
\colhead{RC3} &
\colhead{} &
\colhead{mag} &
\colhead{mag/${\prime\prime}^2$} &
\colhead{kpc} &
\colhead{kpc} &
\colhead{} &
\colhead{\arcdeg} &
\colhead{\arcdeg} &
\colhead{\kms} &
\colhead{$\odot$} &
\colhead{$\odot$} &
\colhead{$\odot$} \\
\colhead{(1)} &
\colhead{(2)} &
\colhead{(3)} &
\colhead{(4)} &
\colhead{(5)} &
\colhead{(6)} &
\colhead{(7)} &
\colhead{(8)} &
\colhead{(9)} &
\colhead{(10)} &
\colhead{(11)} &
\colhead{(12)} &
\colhead{(13)} &
\colhead{(14)} &
\colhead{(15)} &
\colhead{(16)} 
}

\startdata
e383g02 (al) &         &  85.40  &     5  &     I &  12.22 & 19.76 & 16.82 & 4.87 & 0.10 &   60  &   213 &  2.58 & 2.86 & 2.29 & 2.81 \\
\pagebreak
e383g88 (am) &         &  59.51  &     4  &     I &  11.69 & 19.78 & 15.25 & 4.77 & 0.02 &   67  &   274 &  2.55 & 2.75 & 4.00 & 2.77 \\
e435g26 (an) & Antlia  &  40.32  &     5  &    II &  10.13 & 19.67 & 20.33 & 3.65 & 0.06 &   51  &   117 &  2.64 & 2.76 & 1.71 & 2.71 \\
e435g50 (ao) & Antlia  &  40.60  &     5  &    II &  13.53 & 21.34 &  5.55 & 2.23 & 0.00 &   82  &    70 &  2.26 & 2.08 & 0.00 & 2.08 \\
e436g39 (ar) &  Hydra  &  51.25  &     4  &    II &  11.81 & 19.97 & 11.39 & 2.32 & 0.29 &   81  &    82 &  2.56 & 3.62 & 0.00 & 2.80 \\
e437g04 (aq) &  Hydra  &  48.10  &     4  &    II &  11.76 & 19.72 & 10.84 & 2.80 & 0.06 &   63  &   320 &  2.56 & 2.65 & 0.00 & 2.49 \\
e437g30 (ar) &  Hydra  &  54.24  &     4  &     I &  10.89 & 19.57 & 19.05 & 5.28 & 0.12 &   77  &   124 &  2.62 & 3.06 & 0.65 & 2.81 \\
e437g31 (as) &  Hydra  &  56.17  &     7  &     I &  12.99 & 20.37 &  8.89 & 3.12 & 0.02 &   52  &   334 &  2.38 & 3.22 & 0.00 & 3.17 \\
e437g34 (at) &  Hydra  &  54.83  &     3  &    II &  14.25 & 21.73 &  5.32 & 2.91 & 0.04 &   63  &    77 &  2.24 & 4.58 & 0.00 & 4.42 \\
e437g54 (au) &  Hydra  &  50.15  &     3  &     I &  12.94 & 20.35 &  8.24 & 2.84 & 0.11 &   83  &    49 &  2.46 & 5.60 & 0.24 & 5.05 \\
e438g08 (av) &         &  124.3  &    10  &     I &  12.66 & 18.60 & 14.27 & 3.15 & 0.18 &   40  &    89 &  2.50 & 0.90 & 0.40 & 0.82 \\
e438g15 (aw) &         &  49.96  &     4  &    II &  11.38 & 21.06 & 15.16 & 2.87 & 0.10 &   71  &   215 &  2.53 & 2.14 & 2.31 & 2.15 \\
e439g18 (ax) &         & 122.20  &     4  &     I &  12.09 & 19.59 & 20.99 & 5.90 & 0.30 &   43  &   276 &  2.76 & 4.15 & 0.00 & 3.20 \\
e439g20 (ay) &         &  59.84  &     4  &     I &  11.84 & 19.06 & 12.28 & 3.02 & 0.08 &   65  &   281 &  2.64 & 3.46 & 0.00 & 3.21 \\
e441g22 (az) &         &  90.40  &     4  &    II &  11.10 & 20.19 & 27.97 & 7.50 & 0.21 &   73  &   176 &  2.77 & 4.26 & 1.07 & 3.70 \\
e444g21 (ba) &         &  60.68  &    10  &     I &  12.86 & 21.67 & 10.97 & 6.44 & 0.05 &   84 &     64 &  2.36 & 4.50 & 0.57 & 4.32 \\
e444g47 (bb) &         &  62.40  &     6  &     I &  12.82 & 19.63 &  9.51 & 2.71 & 0.00 &   71 &     22 &  2.44 & 1.92 & 0.00 & 1.92 \\
e444g86 (bc) &    K27  &  58.18  &    10  &    II &  11.56 & 19.41 & 12.93 & 3.55 & 0.20 &   78 &    252 &  2.62 & 2.86 & 0.51 & 2.46 \\
e445g15 (bd) &    K27  &  60.34  &    10  &    II &  12.05 & 20.26 & 11.12 & 2.51 & 0.52 &   66 &     56 &  2.58 & 4.02 & 1.14 & 3.04 \\
e445g19 (be) &    K27  &  66.05  &     4  &     I &  11.69 & 19.35 & 15.10 & 4.32 & 0.03 &   67 &     70 &  2.60 & 2.37 & 0.51 & 2.32 \\
e445g35 (bf) &    K27  &  68.02  &     3  &     I &  11.69 & 18.96 & 13.27 & 3.38 & 0.20 &   42 &    175 &  2.76 & 4.27 & 3.44 & 4.13 \\
e445g39 (bf) &         &  61.59  &     3  &     I &  11.17 & 18.65 & 15.48 & 3.58 & 0.03 &   61 &     64 &  2.78 & 3.74 & 1.49 & 3.68 \\
e445g58 (bh) &    K27  &  70.78  &     4  &     I &  11.75 & 19.24 & 15.96 & 4.05 & 0.06 &   63 &    330 &  2.60 & 2.42 & 0.03 & 2.29 \\
e445g81 (bi) &         &  61.12  &     4  &    II &  11.36 & 19.76 & 15.80 & 3.60 & 0.07 &   79 &      3 &  2.67 & 2.74 & 0.00 & 2.55 \\
e446g01 (bj) &         &  98.34  &     4  &     I &  12.07 & 19.63 & 18.96 & 5.27 & 0.28 &   53 &    323 &  2.63 & 2.83 & 1.01 & 2.43 \\
e446g17 (bk) &         &  58.52  &     3  &     I &  11.05 & 20.51 & 18.87 & 6.87 & 0.30 &   54 &    148 &  2.60 & 3.11 & 0.24 & 2.45 \\
e501g01 (bl) &  Hydra  &  55.57  &     7  &     I &  12.64 & 21.12 & 11.00 & 5.06 & 0.06 &   55 &    334 &  2.39 & 4.81 & 0.00 & 4.54 \\
e501g11 (bm) &  Hydra  &  54.94  &     7  &    II &  12.51 & 20.46 & 10.37 & 2.46 & 0.01 &   81 &     94 &  2.41 & 2.61 & 0.00 & 2.57 \\
e501g15 (bn) &  Hydra  &  49.43  &     1  &     I &  10.35 & 19.44 & 19.66 & 5.24 & 0.36 &   60 &    111 &  2.75 & 3.46 & 2.47 & 3.19 \\
e501g68 (bo) &  Hydra  &  45.77  &    10  &    II &  11.93 & 21.27 & 12.01 & 3.99 & 0.19 &   70 &     18 &  2.54 & 4.95 & 3.28 & 4.69 \\
e501g86 (bp) &  Hydra  &  54.28  &     4  &    II &  11.78 & 20.84 & 13.67 & 3.07 & 0.08 &   60 &    201 &  2.52 & 2.75 & 0.49 & 2.59 \\
e502g02 (bq) &  Hydra  &  56.91  &     3  &     I &  11.55 & 18.85 & 12.42 & 2.90 & 0.13 &   63 &     73 &  2.64 & 2.61 & 1.31 & 2.46 \\
e509g80 (br) &         &  92.86  &     4  &    II &  11.86 & 20.70 & 20.29 & 4.36 & 0.14 &   61 &    171 &  2.70 & 3.95 & 2.38 & 3.75 \\
e509g91 (bs) &         &  72.29  &     6  &    II &  12.70 & 21.23 & 13.61 & 3.68 & 0.09 &   79 &    131 &  2.46 & 2.96 & 0.04 & 2.71 \\
e510g11 (bt) &         &  81.13  &     1  &     I &  11.84 & 19.01 & 17.65 & 3.52 & 0.13 &   66 &     64 &  2.68 & 3.02 & 1.88 & 2.89 \\
e569g17 (bu) &         &  57.77  &     3  &     I &  11.99 & 18.29 &  8.03 & 1.69 & 0.05 &   45 &    171 &  2.55 & 1.48 & 0.19 & 1.41 \\
e572g17 (bv) &         &  93.44  &     0  &     I &  11.55 & 18.76 & 19.22 & 4.56 & 0.12 &   48 &     59 &  2.83 & 3.71 & 1.76 & 3.51 \\
\enddata



\end{deluxetable}

\clearpage 

\begin{center} Table 1 {\it Notes} \end{center}
\newcounter{tab1}
\begin{list}{Column \arabic{tab1}--}{\usecounter{tab1}}

\item ESO-Uppsala Catalog number for galaxies in the
sample. The codes for Figs. \ref{fig_massmod} are
also given.
\item Cluster membership. Cen=Centaurus.
\item Distance in megaparsecs assuming Hubble flow with
respect to the cosmic microwave background frame, and H$_\circ$= 75 \kms\ Mpc$^{-1}$.
\item Hubble type in RC3 (de~Vaucouleurs 1991).
\item Freeman type.
\item Total $I$-band magnitude.
\item Disk central surface brightness in $I$.
\item Major axis radius of the $I$ = 23.5 mag arcsec$^{-2}$
isophote in kpc.
\item Exponential disk scale length. We quote the scale
length of the outer disk for Freeman type II galaxies.
\item Bulge to Disk luminosity ratio.
\item Inclination.
\item Major axis position angle measured North through East.
\item log of the circular rotation velocity width in \kms.
\item $I$-band M/L of the disk in solar units.
\item $I$-band M/L of the bulge in solar units.
\item $I$-band M/L of the disk plus bulge in solar units.
\end{list}


\setcounter{figure}{0}

\newpage 
\epsscale{.85}
\clearpage
\begin{figure}
\plotone{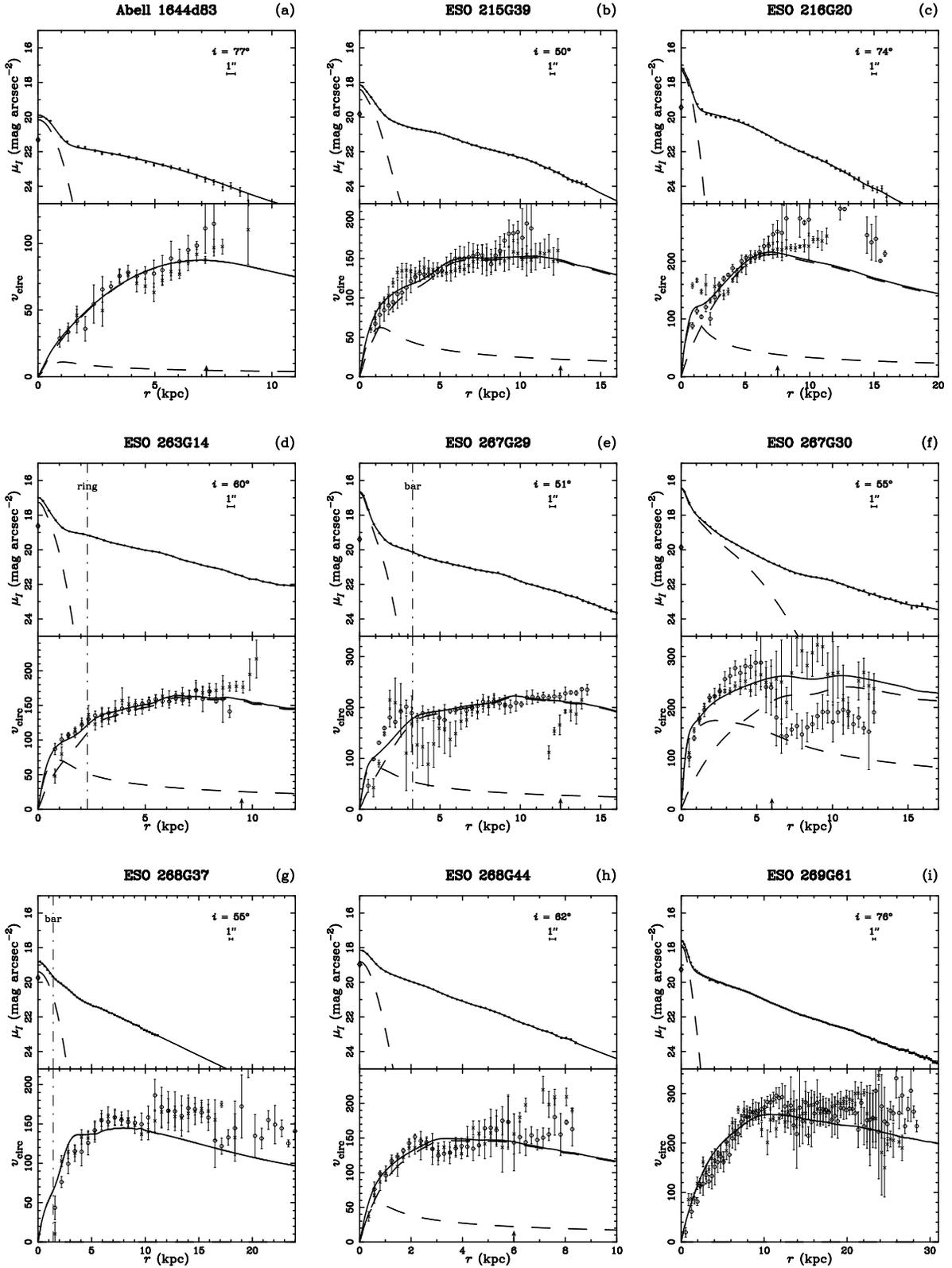}

\caption
{Mass models.  Top panels: $I$-band surface brightness profile. The
bulge profile is given by the dashed line. The diamond indicates the
disk central surface brightness. Lower panels: Maximum disk fits to
the rotation curves. The dashed lines indicate the bulge and disk
contributions. See text for a full description.
\label{fig_massmod}}
\end{figure}

\clearpage
\begin{figure}
\plotone{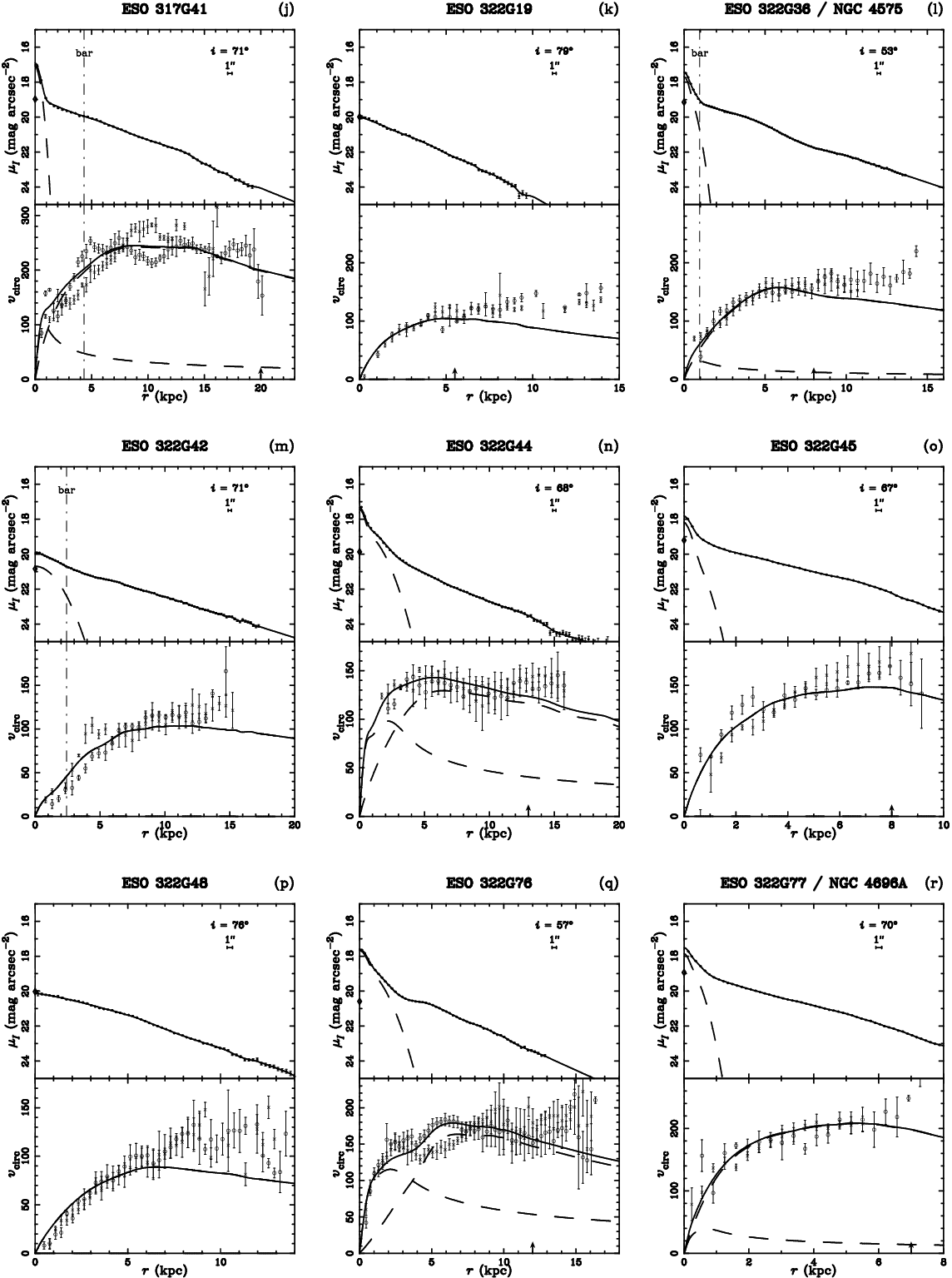}
\setcounter{figure}{0}

\caption
{{\it Continued}}
\end{figure}

\clearpage
\begin{figure}
\plotone{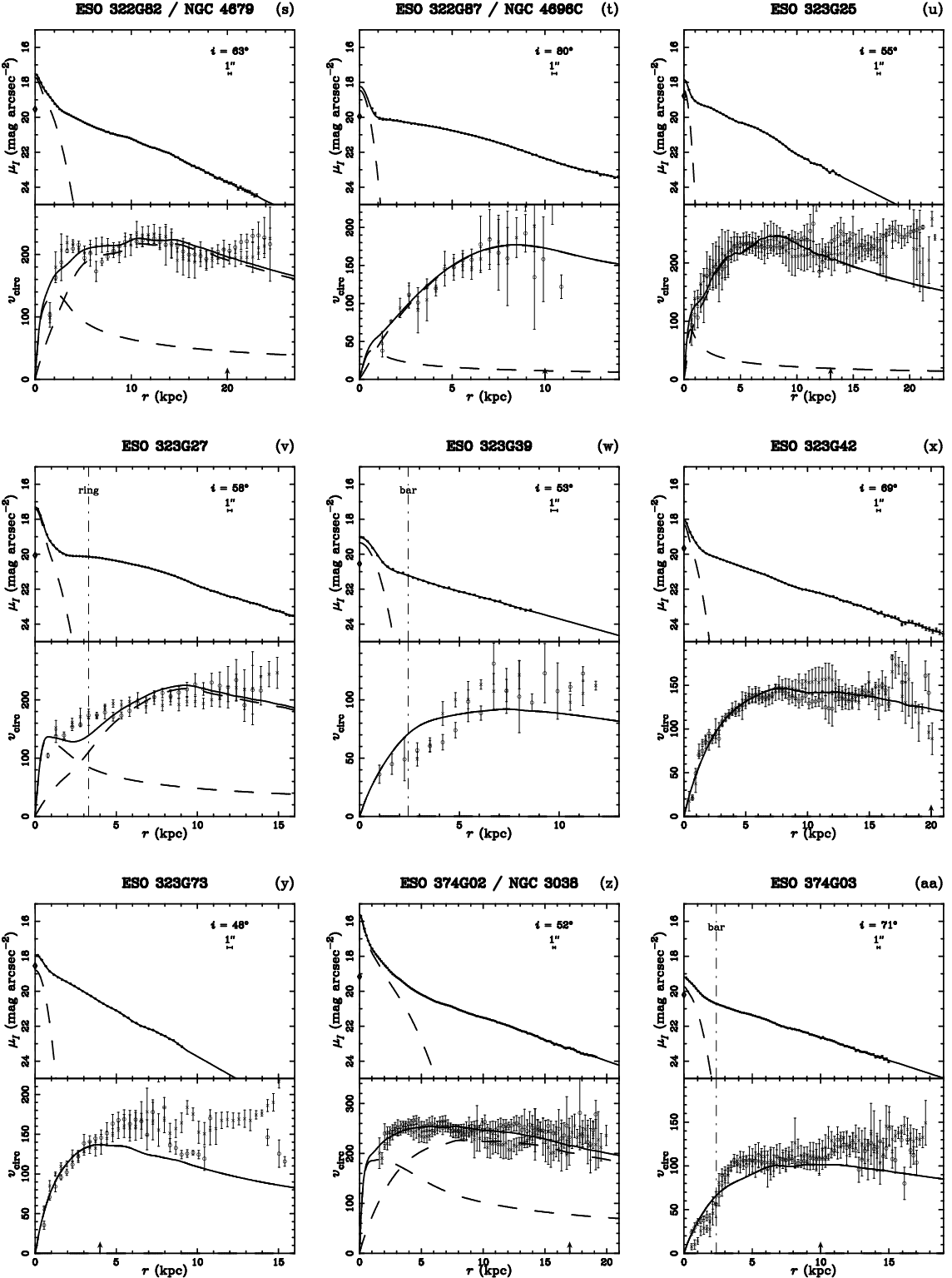}
\setcounter{figure}{0}

\caption
{{\it Continued}}
\end{figure}

\clearpage
\begin{figure}
\plotone{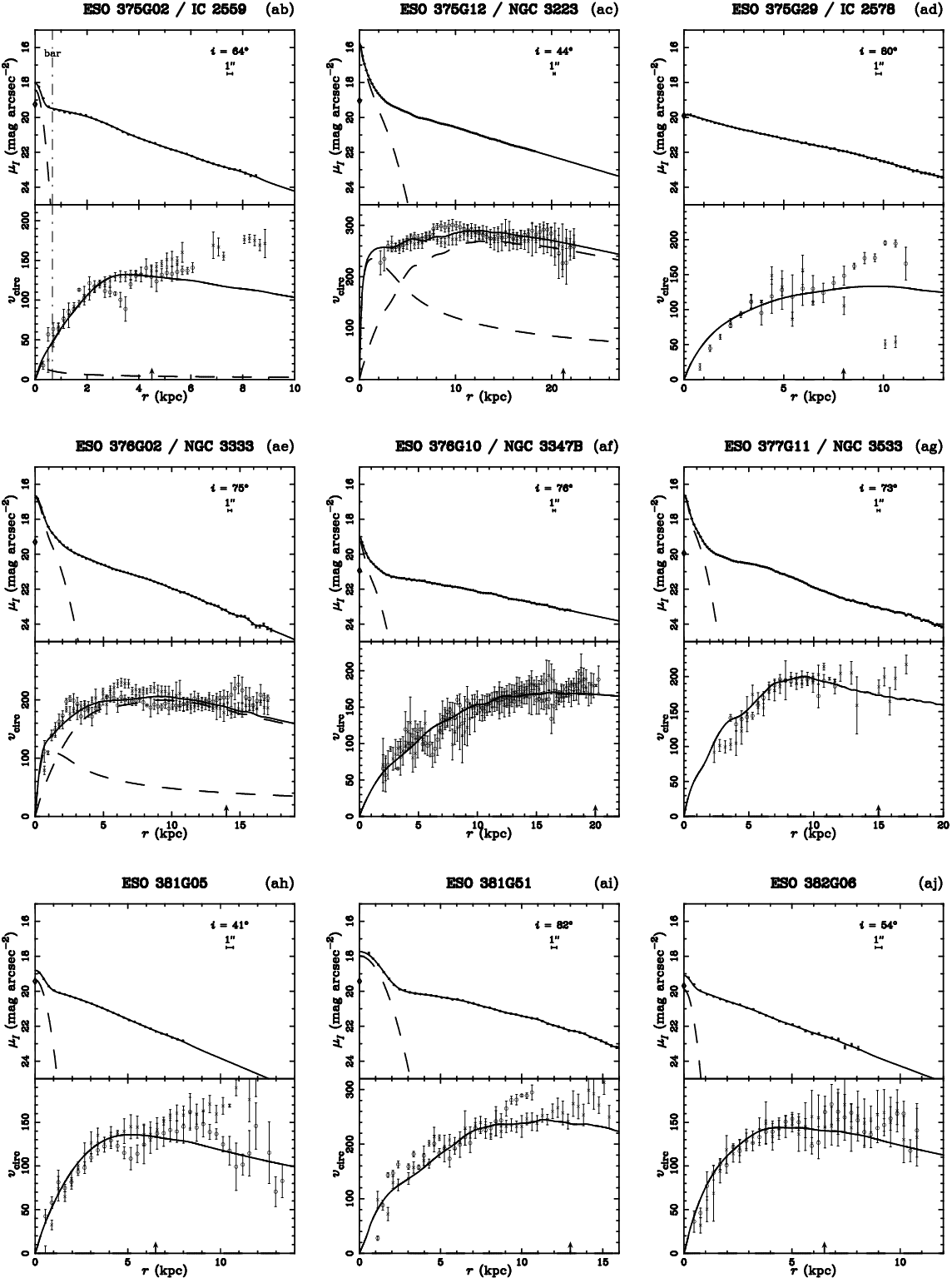}
\setcounter{figure}{0}

\caption
{{\it Continued}}
\end{figure}

\clearpage
\begin{figure}
\plotone{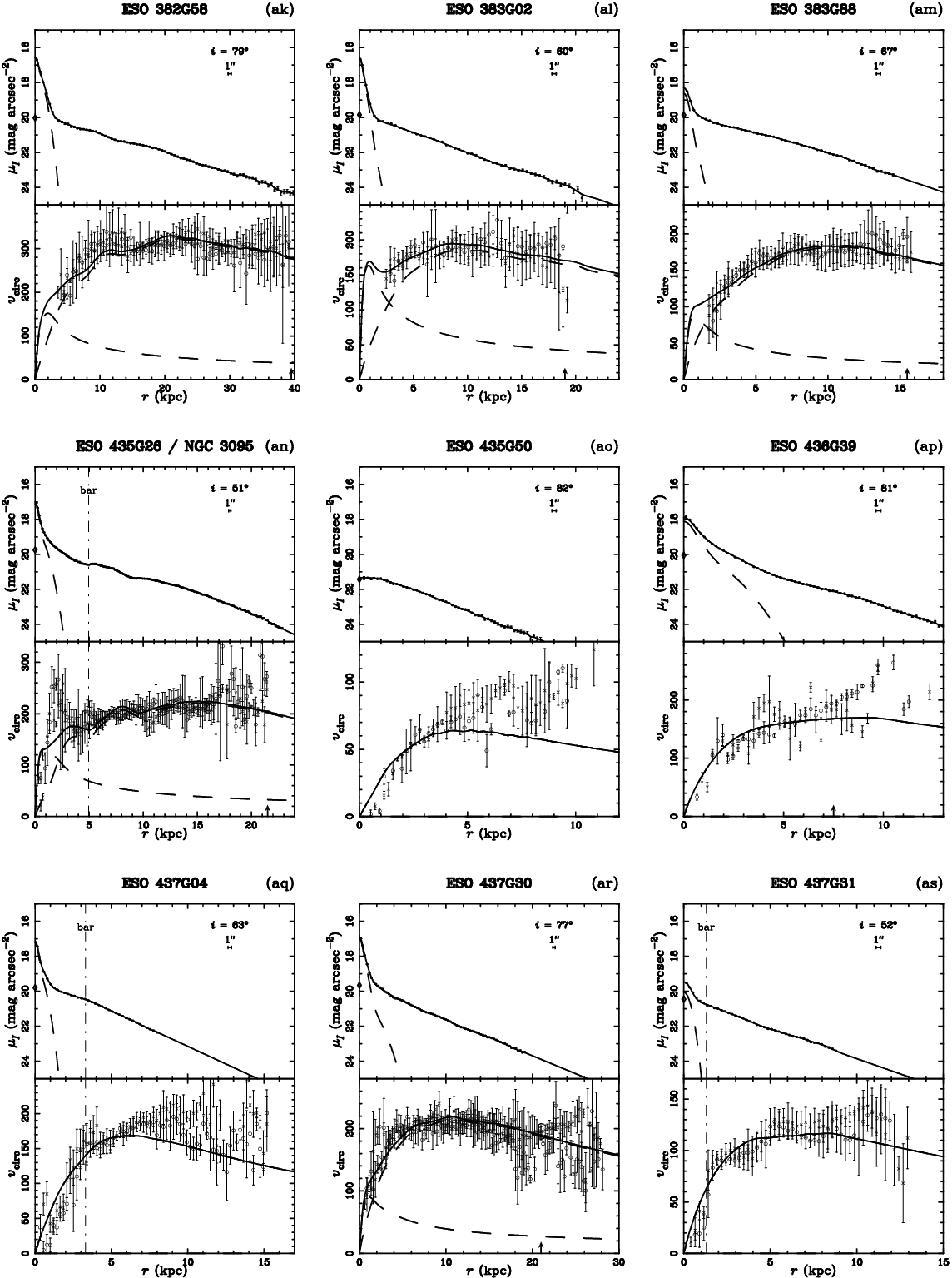}
\setcounter{figure}{0}

\caption
{{\it Continued}}
\end{figure}

\clearpage
\begin{figure}
\plotone{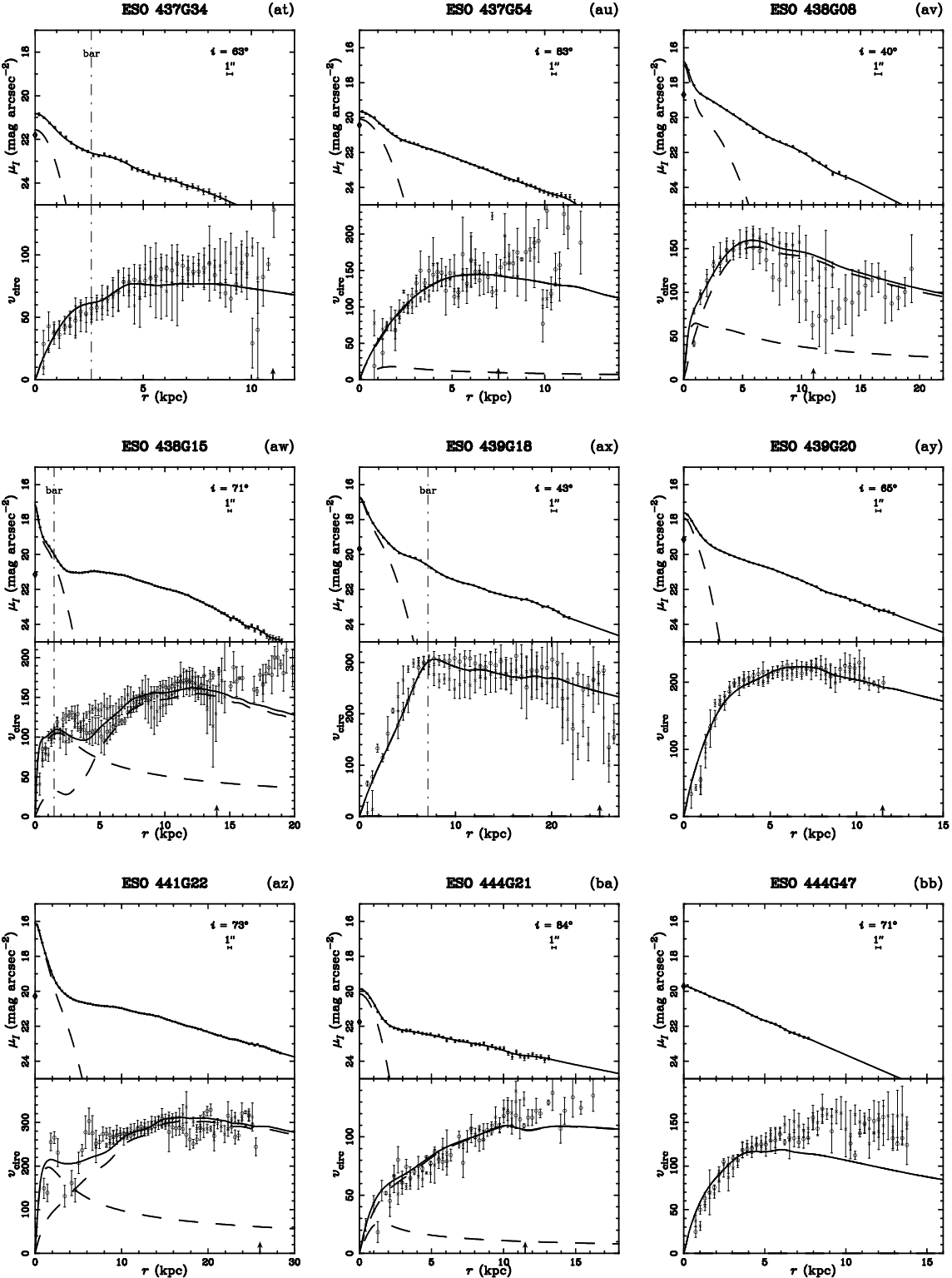}
\setcounter{figure}{0}

\caption
{{\it Continued}}
\end{figure}

\clearpage
\begin{figure}
\plotone{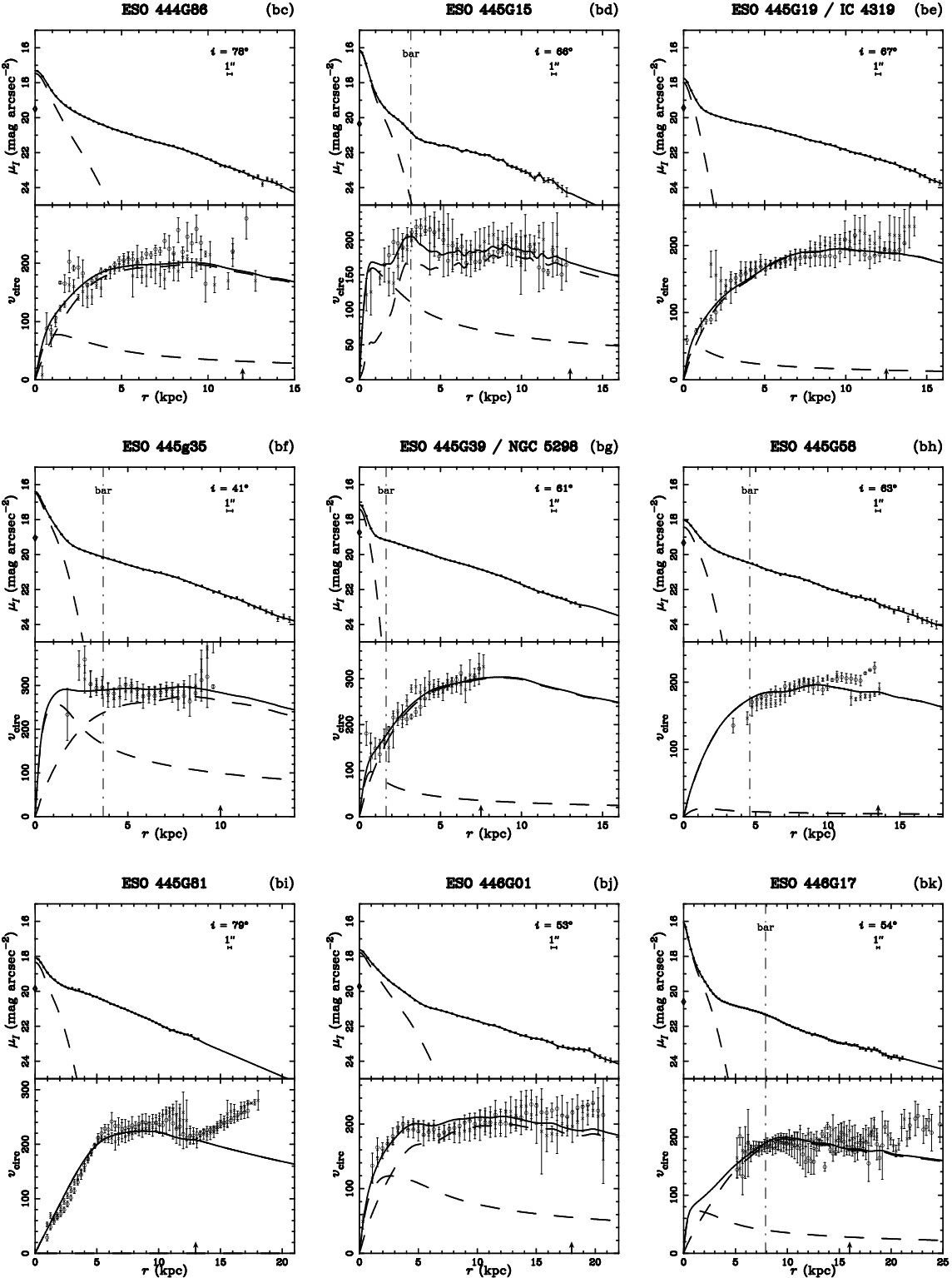}
\setcounter{figure}{0}

\caption
{{\it Continued}}
\end{figure}

\clearpage
\begin{figure}
\plotone{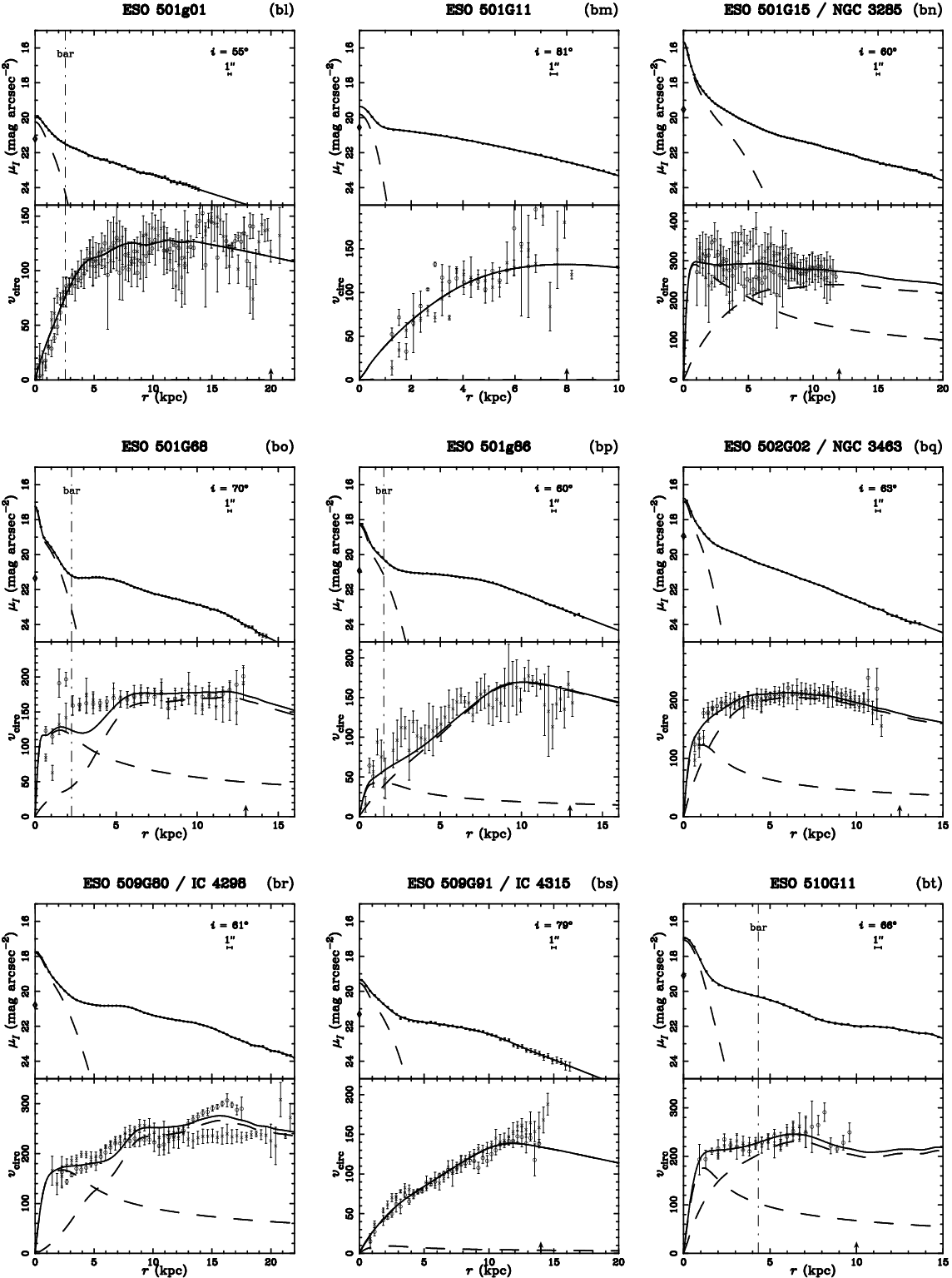}
\setcounter{figure}{0}

\caption
{{\it Continued}}
\end{figure}

\clearpage
\begin{figure}
\plotone{Palunas.fig1_9.ps}
\setcounter{figure}{0}

\caption
{{\it Continued}}
\end{figure}
\epsscale{.8}

\clearpage
\begin{figure}
\plotone{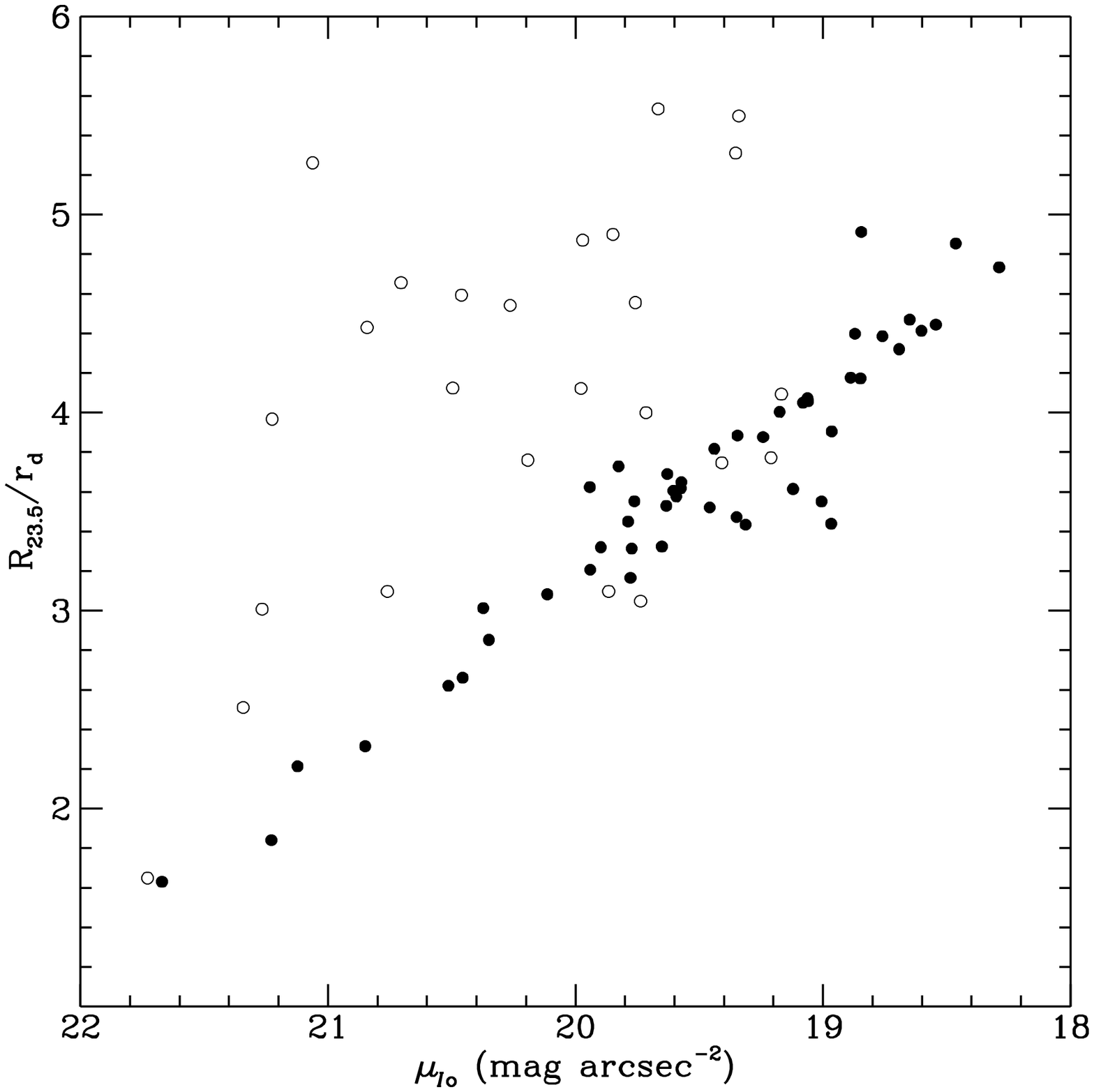}

\caption
{Disk central surface brightness vs.\ \Ropt/r$_{\rm d}$ the solid
symbols are for exponential, Freeman type I disks and the open symbols
are for Freeman Type II disks with a flat central disk profile.
\label{fig_csbR}}
\end{figure}

\clearpage
\begin{figure}
\plotone{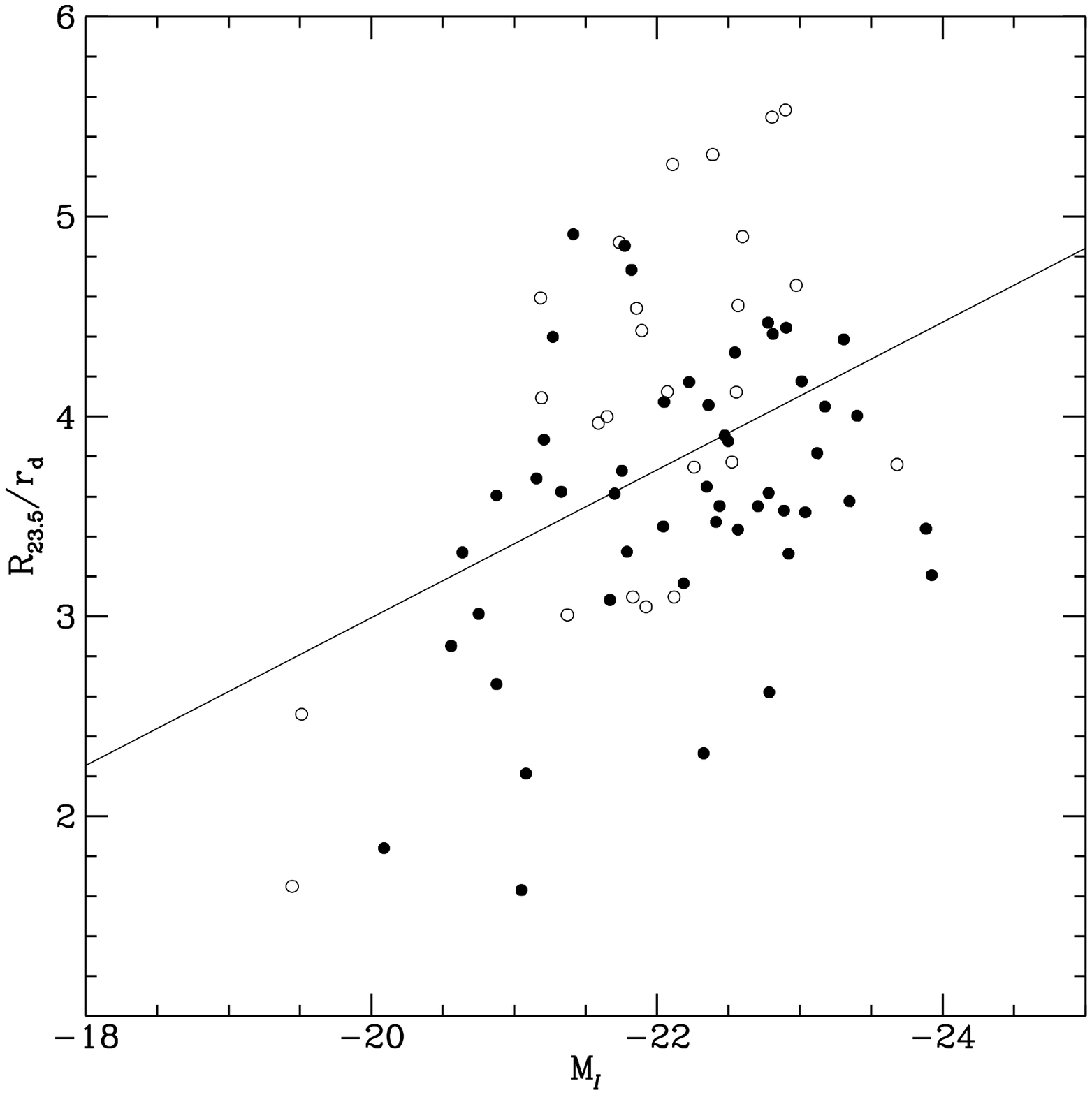}

\caption
{Absolute magnitude vs.\ \Ropt/r$_{\rm d}$.  The symbols are the same
as in Fig.\ \ref{fig_csbR}. The slope of the best fit least squares
line is $-0.37$.
\label{fig_MR}}
\end{figure}

\clearpage
\begin{figure}
\plotone{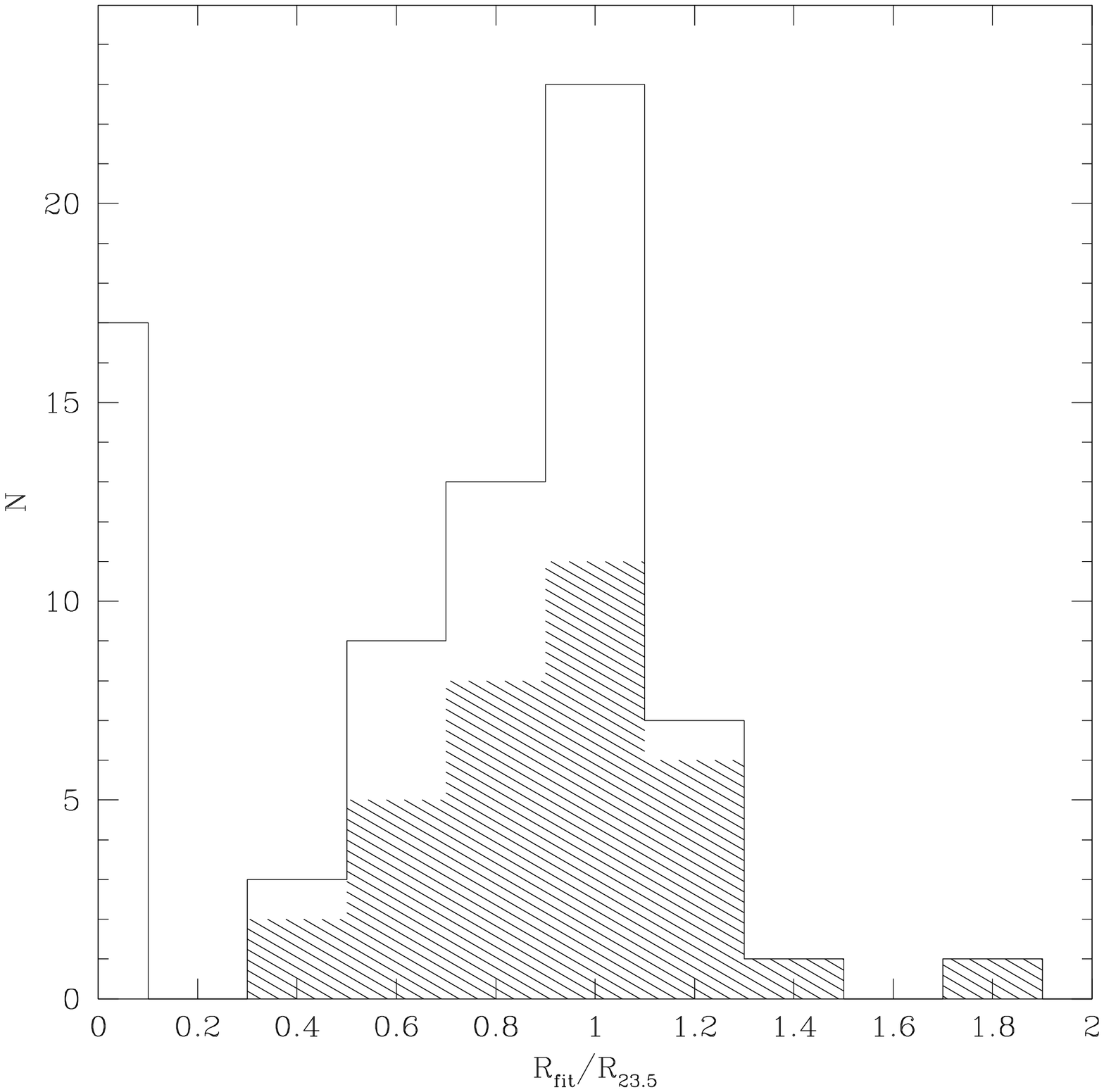}

\caption
{Histogram of R$_{fit}$/\Ropt. The shaded regions indicate galaxies that have
good fits out to the last measured point. Cases where R$_{fit}$/\Ropt = 0
indicate galaxies which have poor fits due to bars or strong spiral arms.
\label{fig_fithist}}
\end{figure}

\clearpage
\begin{figure}
\plotone{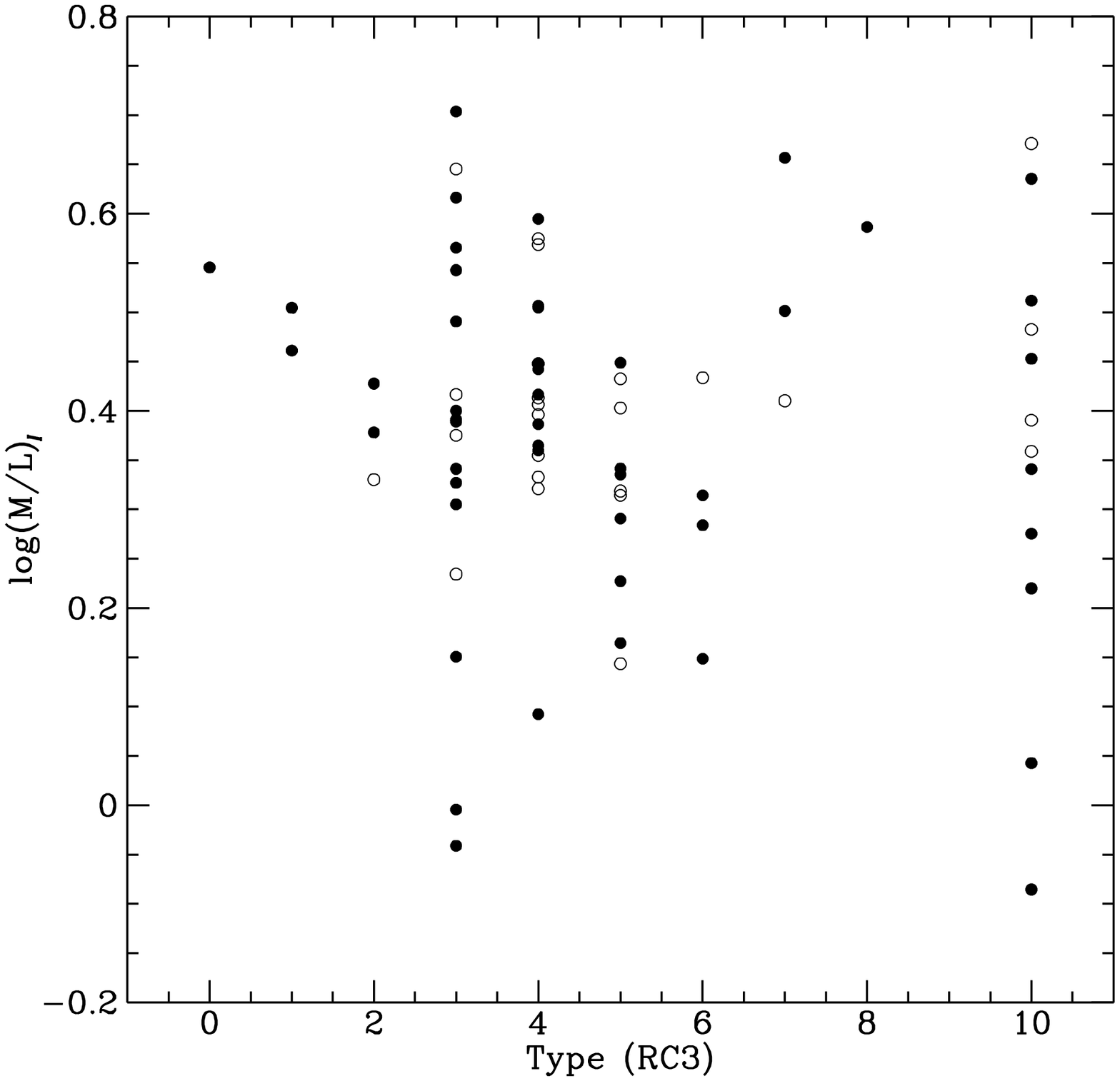}

\caption
{M/L$_I$ vs.\ Hubble type.  The symbols are the same as in
Fig. \ref{fig_csbR}. The Hubble types, classified in the RC3
(de~Vaucouleurs \etal 1991), are coded: Sa=1, Sb=3, Sc=5, Sd=7,
Untyped=10.
\label{fig_mltype}}
\end{figure}

\clearpage
\begin{figure}
\plotone{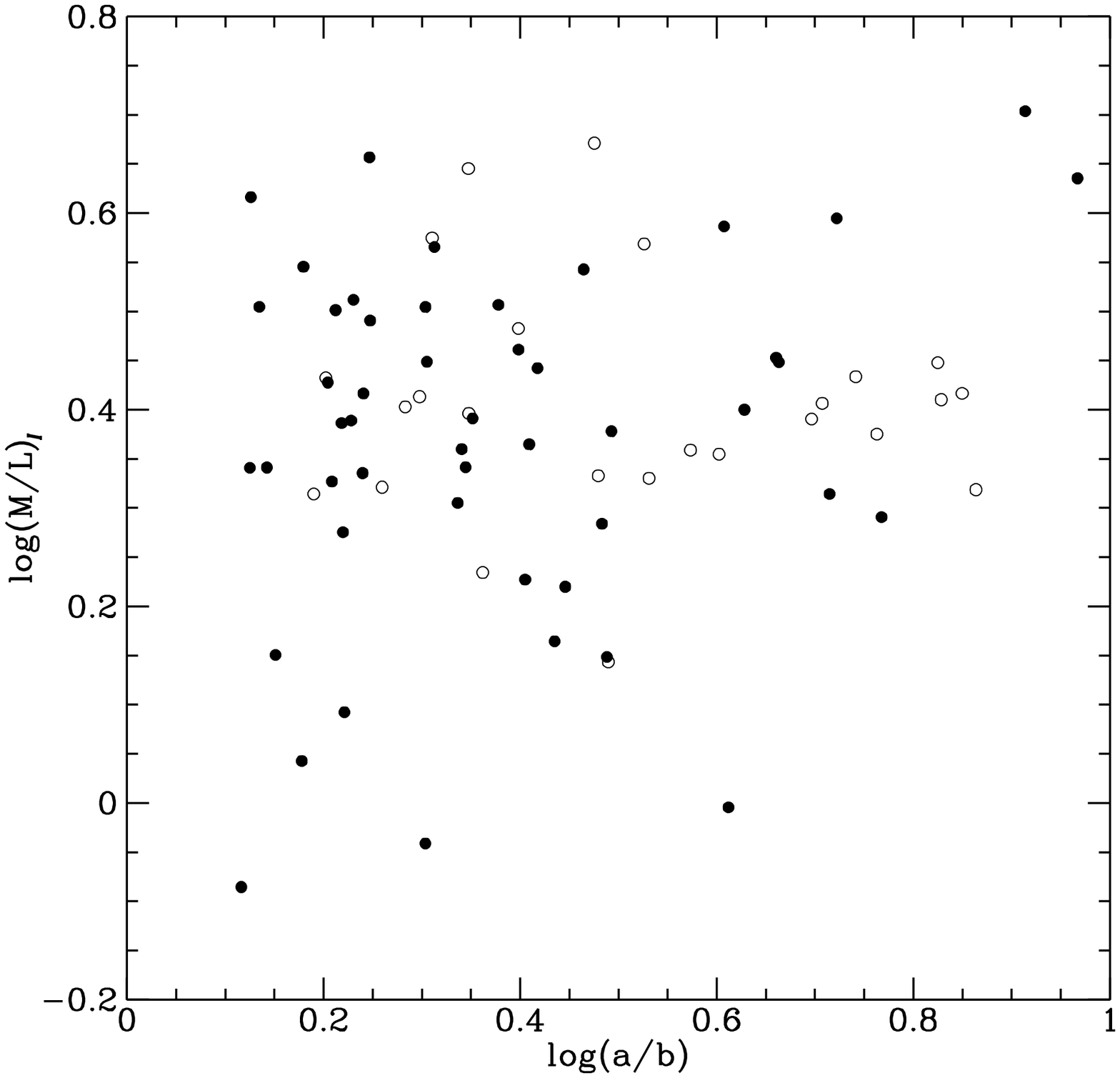}

\caption
{M/L$_I$ vs.\  axis ratio. The symbols are the same as in Fig. \ref{fig_csbR} 
\label{fig_mlax}}
\end{figure}

\clearpage
\begin{figure}
\plotone{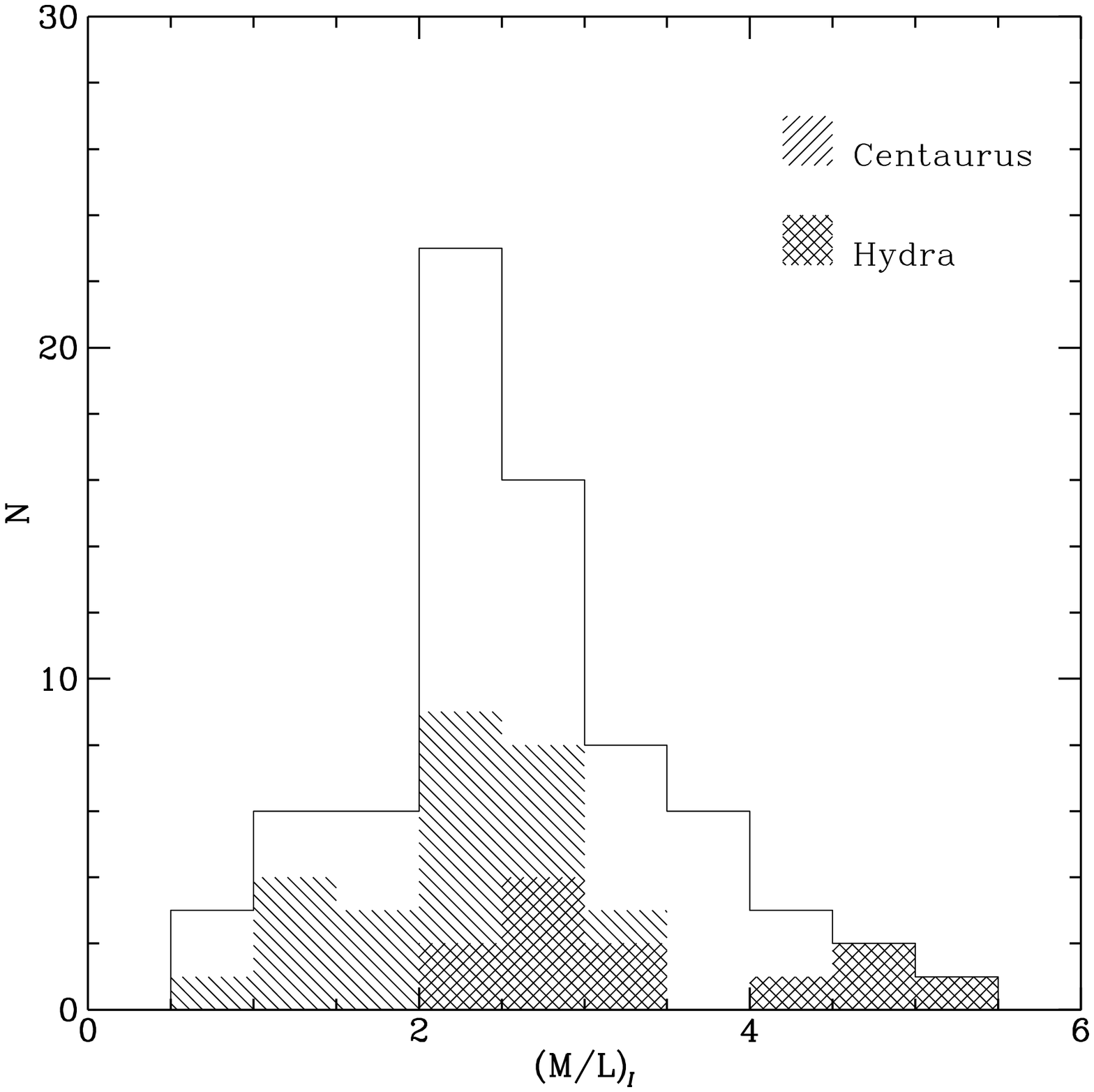}

\caption
{Histogram of M/L$_I$. The hatched regions give the distribution of
M/L in Centaurus and Hydra clusters.
\label{fig_mlhist}}
\end{figure}

\clearpage
\begin{figure}
\plotone{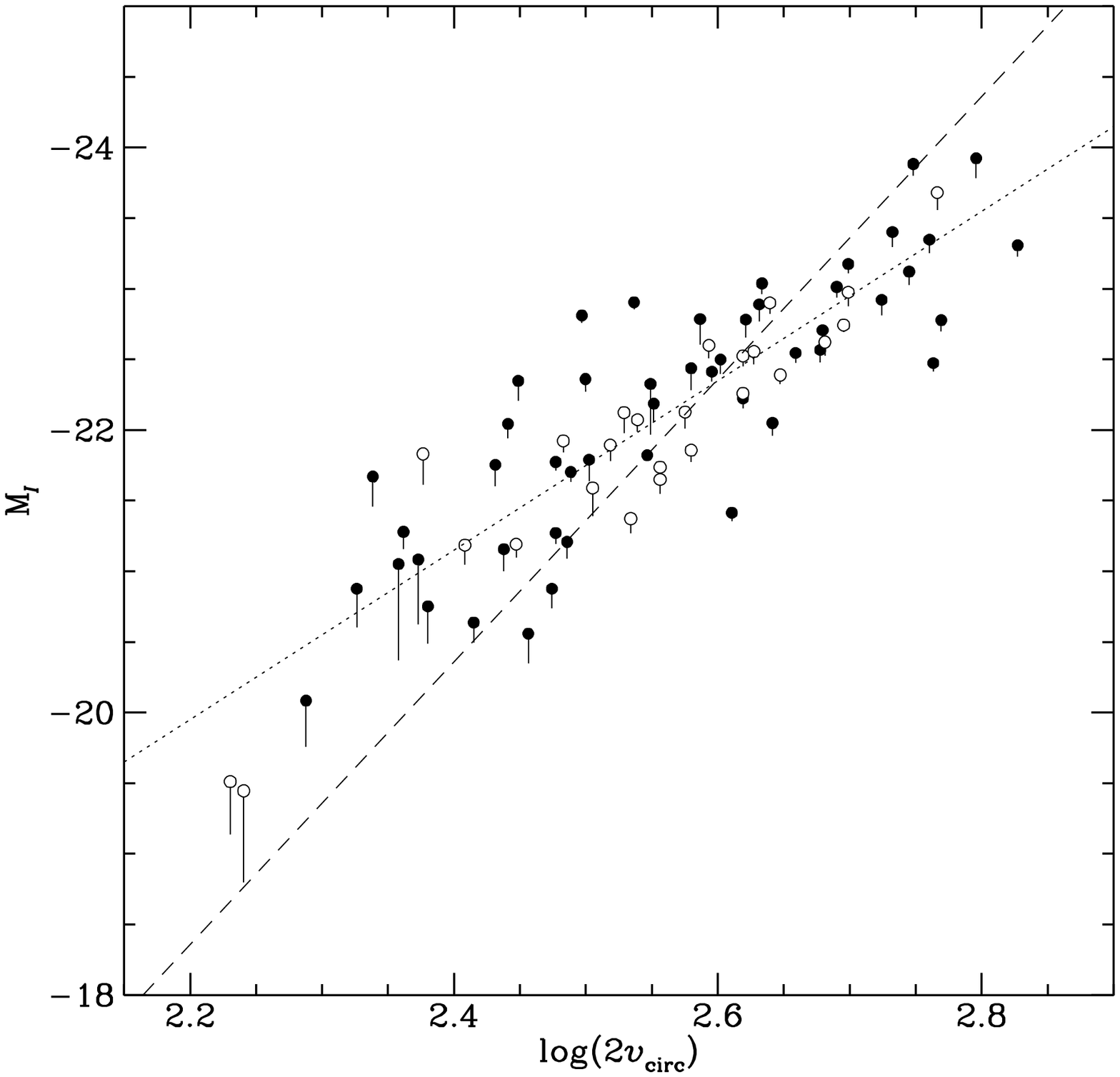}

\caption
{The Tully-Fisher relation.  The symbols are the same as in Fig.\
\ref{fig_csbR}. The symbols mark the total extrapolated magnitude and
the lower points of the vertical lines mark the \Ropt\ isophotal
magnitudes. The dashed line is the TF relation assuming a slope of 10,
and the dotted line is the TF relation assuming a slope of 6.
\label{fig_TF}}
\end{figure}

\end{document}